\documentclass[
amsmath,amssymb,
aps,
notitlepage,
prd
]{revtex4-1}

\usepackage{graphicx}
\usepackage{dcolumn}
\usepackage{bm}

\usepackage{tensor}

\usepackage{xcolor}

\usepackage{hyperref}

\begin{document}

\title{ 
Quasinormal modes of dilatonic Reissner-Nordstr\"om black holes
} 
 
\author{Jose Luis Bl\'azquez-Salcedo}
\email[]{jose.blazquez.salcedo@uni-oldenburg.de}
\author{Sarah Kahlen}
\email[]{sarah.kahlen1@uni-oldenburg.de}
\author{Jutta Kunz}
\email[]{jutta.kunz@uni-oldenburg.de}
\affiliation{Institut f\"ur Physik, Universit\"at Oldenburg, D-26111 Oldenburg, Germany}

\date{\today}

\begin{abstract}
We calculate the quasinormal modes of static spherically symmetric 
dilatonic Reissner-Nordstr\"om black holes for general values of 
the electric charge and of the dilaton coupling constant. 
The spectrum of quasinormal modes is composed of five families of modes: 
polar and axial gravitational-led modes, 
polar and axial electromagnetic-led modes,
and polar scalar-led modes. 
We make a quantitative analysis of the spectrum, revealing its dependence 
on the electric charge and on the dilaton coupling constant.
For large electric charge and large dilaton coupling, 
strong deviations from the Reissner-Nordstr\"om modes arise. 
In particular, isospectrality is strongly broken,
both for the electromagnetic-led and the gravitational-led modes,
for large values of the charge.
\end{abstract}


\maketitle

\section{Introduction}

In recent years 
the LIGO-VIRGO collaboration has reported the direct detection of 
gravitational waves from merging black holes 
\cite{PhysRevLett.116.061102, Abbott:2016nmj,PhysRevLett.119.141101, PhysRevLett.118.221101,LIGOScientific:2018mvr}, 
and neutron stars \cite{Abbott:2017xzg}. 
These detections have also been studied in the electromagnetic spectrum 
\cite{Coulter:2017wya, TheLIGOScientific:2017qsa, GBM:2017lvd,Abbott:2018wiz}, 
representing new examples of multi-messenger astronomy \cite{Neronov:2019uht}. 
Moreover, the current O3 run of the LIGO-VIRGO collaboration 
is already reporting a large amount of new events \cite{GraceDB}. 

Gravitational waves following the merging of black holes possess 
a ringdown phase characterized by a spectrum of frequencies and damping times. 
This spectrum can be studied using quasinormal modes (QNMs) 
(see e.g.~\cite{Kokkotas:1999bd,Nollert:1999ji,Berti:2009kk,Konoplya:2011qq}). 
Based on the next generations of gravitational wave detectors,
it will become possible to directly test the regime of strong gravity 
by comparing the theoretically predicted ringdown spectrum 
of black holes with direct measurements 
(see e.g.~\cite{Berti:2015itd,Berti:2018vdi,Barack:2018yly}).

One interesting aspect here is to test for the existence of scalar hair 
on black holes \cite{Isi:2019aib}. 
In General Relativity (GR) coupled to Maxwell electrodynamics, 
i.e., in Einstein-Maxwell (EM) theory,
black holes possess no hair. They are uniquely
described by their global charges, mass $M$ and angular momentum $J$,
and electric $Q$ and magnetic charge $P$
(as discussed in detail, e.g. in \cite{Chrusciel:2012jk}).
However, there are various mechanisms that allow to circumvent
the no-hair theorem \cite{Herdeiro:2015waa,Cardoso:2016a}, 
leading to scalar hair on black holes. 
These typically involve non-trivial couplings of a scalar field
to an invariant. 

From a string theory perspective, the scalar field would correspond
to a dilaton, with an exponential coupling to the invariant.
Choosing the invariant to be the
Lagrangian of the electromagnetic field one obtains
the static dilatonic charged black holes of Gibbons and Maeda
\cite{Gibbons:1987ps,Garfinkle:1990qj}
and their rotating generalizations
\cite{Frolov:1987rj,Rasheed:1995zv,Kleihaus:2003df}.
Choosing for the invariant a curvature invariant,
as for instance, the Gau\ss-Bonnet term, 
Einstein-Gau\ss -Bonnet-dilaton (EGBd) black holes emerge
\cite{Kanti:1995vq,Kleihaus:2011tg,Blazquez-Salcedo:2016yka,Kleihaus:2015aje,Kokkotas:2017ymc}.
In both cases the black holes carry non-trivial scalar hair,
and the Reissner-Nordstr\"om (RN) and Kerr black holes, respectively,
are no longer solutions of the field equations.
We note that in both cases recently, 
much interest has focused on more general coupling functions
(see e.g.~\cite{Doneva:2017bvd,Silva:2017uqg,Antoniou:2017acq,Herdeiro:2018wub,Cunha:2019dwb,Konoplya:2019goy}),
since these allow for spontaneously scalarized black holes,
when the leading term of the coupling function is quadratic 
in the scalar field.

Here we will focus on static spherically symmetric
charged black holes with dilatonic coupling
function to the Lagrangian of the electromagnetic field.
The resulting Einstein-Maxwell-dilaton (EMD) theory then features
a dilaton coupling constant $\gamma$ in the coupling function,
which we treat as a free parameter, following Gibbons and Maeda
\cite{Gibbons:1987ps,Dobiasch:1981vh,Garfinkle:1990qj}.
We note that for $\gamma=0$, the dilaton decouples
and the RN black holes are recovered.
Interesting non-trivial special cases represent $\gamma=1$,
leading to the static charged string theory black holes
studied also by Garfinkle-Horowitz-Strominger (GHS) \cite{Garfinkle:1990qj},
and $\gamma=\sqrt{3}$, yielding the Kaluza-Klein (KK) black holes
\cite{Dobiasch:1981vh,Gibbons:1987ps}.

The presence of the dilaton has profound consequences for the properties
of the black holes. Not only do they carry scalar hair, but their domain
of existence undergoes a fundamental change with respect to the EM case.
The extremal RN solutions possess a maximal charge
with $Q/M=1$ and a finite horizon area. However, when
the static electrically charged EMD black holes approach extremality,
their maximal possible charge %
exceeds the RN value the more the larger the coupling constant, %
while the horizon area of the limiting solution tends to zero 
\cite{Gibbons:1987ps}.

Like the RN black holes, the dilatonic black holes emerge from the 
Schwarzschild black holes when the electric charge is increased
from zero. The RN black holes are (mode)-stable 
under linear perturbations, as evaluation of their QNMs has shown
\cite{Moncrief:1974gw,Moncrief:1974ng,Moncrief:1975sb,Gunter:1980,Kokkotas:1988fm,Leaver:1990zz,Andersson:1996xw,Richartz:2014jla}.
QNMs of the dilatonic GHS ($\gamma=1$) black holes have been studied in 
\cite{Ferrari:2000ep}, 
while axial QNMs of dilatonic black holes were investigated for
general values of the dilaton coupling $\gamma$
in \cite{Konoplya:2001ji}.
Recently, the analysis was extended to polar QNMs as well, but 
restricting to small values of the charge of the EMD black holes
\cite{Brito:2018hjh}.
Similarly, QNMs of EGBd black holes have received much interest
\cite{Pani:2009wy,Blazquez-Salcedo:2016enn,Blazquez-Salcedo:2017txk,Konoplya:2019hml,Zinhailo:2019rwd},
and likewise QNMs of spontaneously scalarized black holes 
\cite{Blazquez-Salcedo:2018jnn,Silva:2018qhn,Myung:2018jvi,Myung:2018vug,Myung:2019oua}.

The linear perturbations can be split into axial and polar perturbations,
according to their transformation 
under the reflection of the angular coordinates. 
As shown in \cite{Ferrari:2000ep} for a dilatonic black hole 
the excitation of axial modes leads to the 
simultaneous emission of gravitational and electromagnetic waves, 
whereas the excitation of polar modes includes in addition the emission 
of scalar radiation. 
Whereas for Schwarzschild and RN black holes 
axial and polar gravitational waves are emitted with the same frequencies, 
since the corresponding potentials of the wave equation are related, 
leading to the same reflection and transmission coefficients, 
this isospectrality is broken for the dilatonic black holes.

Our objective here is to investigate the spectrum of QNMs 
of charged dilatonic black holes for general coupling constant $\gamma$,
allowing for any value of the electric charge up to the 
(respective) maximal charge.
In Section II we define the theory,
introduce the ansatz for static spherically symmetric black holes,
and recall some of their properties. Subsequently in Section III, 
linear perturbations are introduced and discussed 
for three different cases:
purely spherical, axial and polar perturbations. 
Since in the general case of charged dilatonic black holes 
gravitational, electromagnetic and scalar perturbations become coupled, 
the QNMs are further categorized into gravitational-led, 
electromagnetic-led and scalar-led perturbations. 

We present our numerical results in Section IV
for $0\le l \le 2$ and several values 
of the dilaton coupling constant $\gamma$. 
We compare related spectra of these three categories 
as well as polar and axial perturbations.
Our results confirm that the isospectrality of the electro-vac black holes
is broken in the presence of a dilaton. 
In the limit of vanishing charge the Schwarzschild QNMs are recovered.  
Likewise, for dilaton coupling $\gamma=0$
and $\gamma=1$, the RN and GHS QNMs are regained.

\section{Black holes in Einstein-Maxwell-dilaton theory}

\subsection{Theory and ansatz}

We consider the EMD action
\begin{eqnarray}
I
=\frac{1}{2\kappa}\int_{\mathcal{M}}
{d^{4}x\sqrt{-g}}\left[
R-e^{\gamma\phi}F_{\mu\nu}F^{\mu\nu}
-\frac{1}{2}\partial_\mu\phi\,\partial^\mu\phi-V(\phi)\right] \ ,
\label{action1}
\end{eqnarray}
where $\kappa$ is the gravitational constant, $R$ is the Ricci scalar, 
$F_{\mu\nu}=\partial_{\mu}A_\nu -\partial_{\nu}A_\mu $ 
is the Maxwell field and $\phi$ is the dilaton field. 
The dilaton field is coupled to the electromagnetic field 
via an exponential coupling $e^{\gamma\phi}$, 
where $\gamma$ is the dilaton coupling constant. 
The scalar field may be supplemented with a potential $V(\phi)$. 

The resulting field equations are
\begin{eqnarray}
\label{Eins} 
R_{\mu\nu}-\frac{1}{2}g_{\mu\nu}R=
T_{\mu\nu}^{\phi}+T_{\mu\nu}^{EM} \ , \\
\nabla_{\mu}(\sqrt{-g}e^{\gamma\phi}F^{\mu\nu} )
=0 \ , \label{Mxw}
\\
\frac{1}{\sqrt{-g}}\partial_{\mu}
 (\sqrt{-g}g^{\mu\nu}\partial_{\nu}\phi )
 =\frac{dV(\phi)}{d\phi}+\gamma{} {e}^{\gamma\phi}F_{\mu\nu}F^{\mu\nu} \ ,
\label{Klein}
\end{eqnarray}
where we have introduced the dilaton stress energy momentum 
$T_{\mu\nu}^{\phi}$ and the electromagnetic stress-energy-momentum 
$T_{\mu\nu}^{EM}$
\begin{eqnarray}
T_{\mu\nu}^{\phi}
\equiv\frac{1}{2}\partial_{\mu}\phi\partial_{\nu}\phi
-\frac{1}{2}g_{\mu\nu}
 \left( \frac{1}{2}(\partial \phi)^2+V(\phi)  \right) \ , \\
T_{\mu\nu}^{EM}\equiv2e^{\gamma\phi}
\left(F_{\mu\alpha}F_{\nu}^{\,\,\alpha}
-\frac{1}{4}g_{\mu\nu}F^2\right)\ .
\end{eqnarray}
Here we will focus on the case of vanishing dilaton potential
\begin{eqnarray}
V(\phi) = 0.
\end{eqnarray}

The static spherically symmetric EMD black hole solutions 
can be obtained with the line element 
\begin{eqnarray}
\label{stab1}
ds^2=-f(r) dt^2 + \frac{dr^2}{1-2m(r)/r} +r^2(d\theta^2+\sin^2 \theta d\varphi^2) \ ,
\end{eqnarray}
with the metric functions $f$ and $m$. 
The matter fields for the static spherically symmetric solutions
are parametrized by
\begin{eqnarray}
\label{stab2}
A&=&  a_0(r) dt \ , \nonumber \\ \phi&=&\phi_0(r) \ , 
\end{eqnarray} 
where $a_0$ and $\phi_0$ are the electric and the dilaton function,
respectively. 
With this ansatz for the metric and matter fields, 
one obtains the following set of ordinary differential equations 
for $f,m,a_{0}$ and $\phi_{0}$ 
from the field equations (\ref{Eins}), (\ref{Mxw}) and (\ref{Klein}):
\begin{eqnarray}
\label{stateq}
\partial_r\delta &=&-\frac{r}{4}\, \left( \partial_r\phi_{0}  \right)^{2} \ , \nonumber \\ 
\partial_rm &=&\frac{1}{2}\,{{e}^{\gamma\phi_{0} + 2 \delta}} {r}^{2} \left( \partial_ra_0  \right)^{2}+\frac{r}{8} \left(r-2m\right)  \left(\partial_r\phi_{0}\right)^{2} \ , \nonumber \\
\partial^2_r\phi_{0} &=& {\frac {{{e}^{\gamma\,\phi_{0} + 2\delta}} {r}}{r-2m }} \left( r(\partial_r\phi_{0})-2\gamma\right)\left( \partial_r a_0\right)^{2}  \nonumber \\ 
& & +2{\frac{m-r}{\left(r-2m\right)r} (\partial_r\phi_{0})} \ , \nonumber \\
\partial^2_r a_0 &=& \left(\frac{1}{4}\, \left(\partial_r\phi_{0}\right) ^{2}r-\gamma(\partial_r\phi_{0})-\frac{2}{r} \right) (\partial_r a_0) \ , 
\end{eqnarray}
where we have defined $f=\left(1-\frac{2m}{r}\right)e^{-2\delta}$.

The first integral of the electromagnetic field yields
\begin{eqnarray}
\partial_r a_0 = \frac{Q}{e^{\gamma \phi_0 + \delta}r^2} \ ,
\end{eqnarray}
where $Q$ is the electric charge of the configuration. 
This equation can be used to simplify the previous system of equations.

\subsection{Properties of dilatonic Reissner-Nordstr\"om black holes}

The static spherically symmetric electrically charged
EMD black hole solutions have been obtained in closed form
for arbitrary dilaton coupling constant $\gamma$
by Gibbons and Maeda \cite{Gibbons:1987ps}.
In the limit $\gamma=0$, the dilaton field becomes trivial 
and the RN black hole is recovered: 
\begin{eqnarray}
f=1-\frac{2M}{r}+\frac{Q^2}{r^2}  \ , \
m=M-\frac{Q^2}{2r} \ , \
a_0 = -\frac{Q}{r} \ , \
\phi_0 = 0.
\label{RN_sol}
\end{eqnarray} 
In the case of coupling constant $\gamma=1$, 
the stringy GHS black holes are recovered \cite{Garfinkle:1990qj},
which can be uplifted to $N=4$ supergravity. 
When $\gamma=\sqrt{3}$, charged four dimensional Kaluza-Klein black holes
arise from the compactification of five dimensional vacuum black holes 
\cite{Dobiasch:1981vh}.
When besides electric charge $Q$, 
also magnetic charge $P$ is present, dyonic black holes result
\cite{Dobiasch:1981vh,Gibbons:1987ps,Garfinkle:1990qj,Astefanesei:2019pfq}.
However, here we will focus on purely electric black holes ($P=0$).

Studying perturbatively the asymptotic behavior 
of the black hole solutions for $r \to \infty$,
we find that asymptotically flat configurations satisfy
\begin{eqnarray}
\label{inf_static}
f &=&1-\frac{2M}{r}+ O(r^{-2}) \ , \nonumber \\
m &=&M-\left(Q^2+Q_S^2/4\right)\frac {1}{2r}+O(r^{-2}) \ , \nonumber \\
\phi_{0} &=&\frac{Q_S}{r} + \left(M Q_S - \gamma Q^2\right)\frac{1}{r^2} + O(r^{-3}) \ , \nonumber \\
a_0 &=&-\frac{Q}{r} + \frac{\gamma Q Q_S}{2r^2} + O(r^{-3}) \ ,
\end{eqnarray}
where $M$ is the total mass of the black hole, 
$Q$ is the electric charge and $Q_S$ the scalar (dilaton) charge. 
Although asymptotically, there are three parameters, 
one of them is not a free parameter \cite{Pacilio:2018gom}. 
For instance, since there is no conservation law 
for the dilaton field \cite{Herdeiro:2015waa}, 
the existence of a horizon imposes a non-trivial relation $Q_S = Q_S(M,Q)$, 
meaning that the dilatonic black hole has secondary scalar hair.

\begin{figure}
	\centering
	\includegraphics[width=0.35\linewidth,angle=-90]{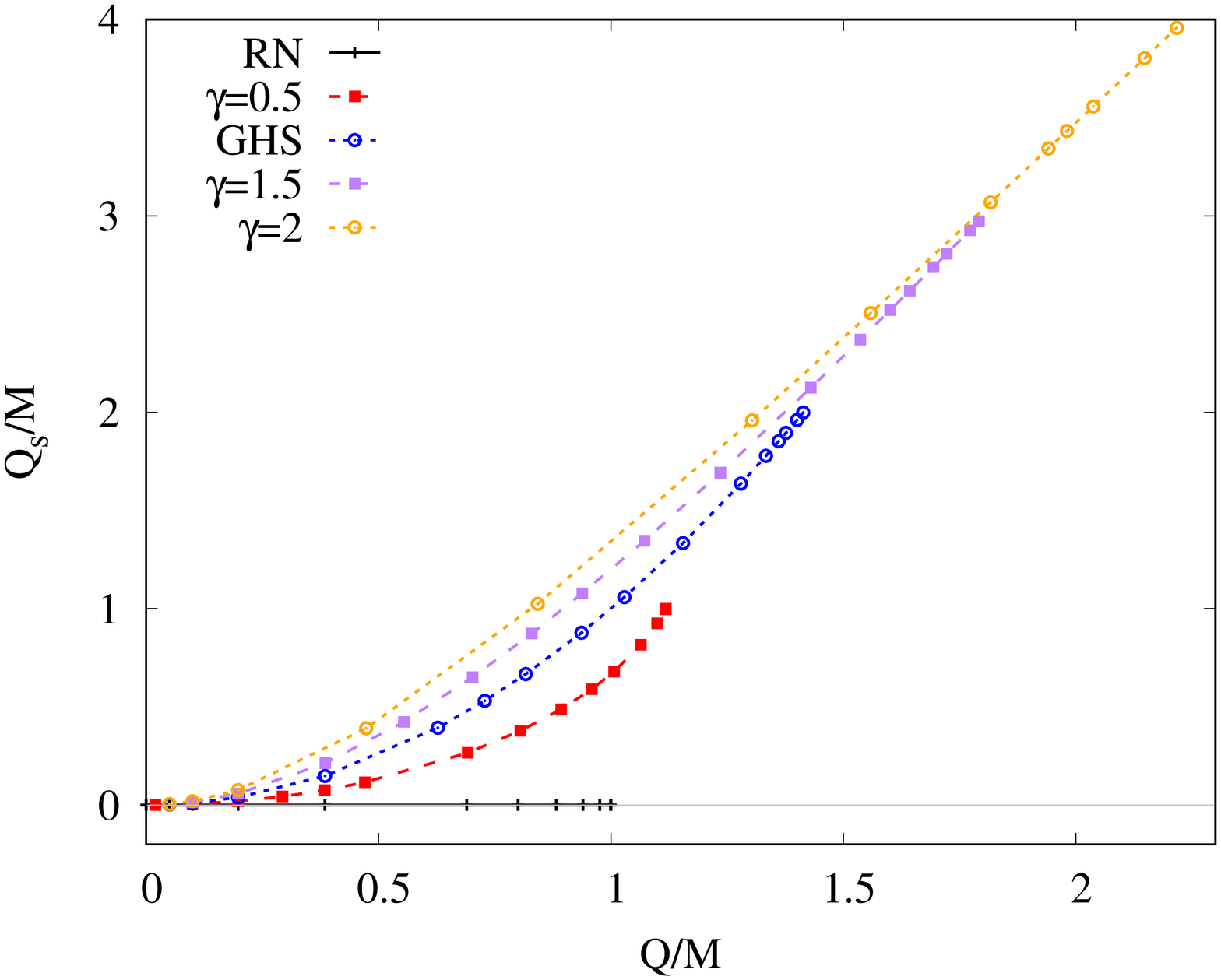}
	\includegraphics[width=0.35\linewidth,angle=-90]{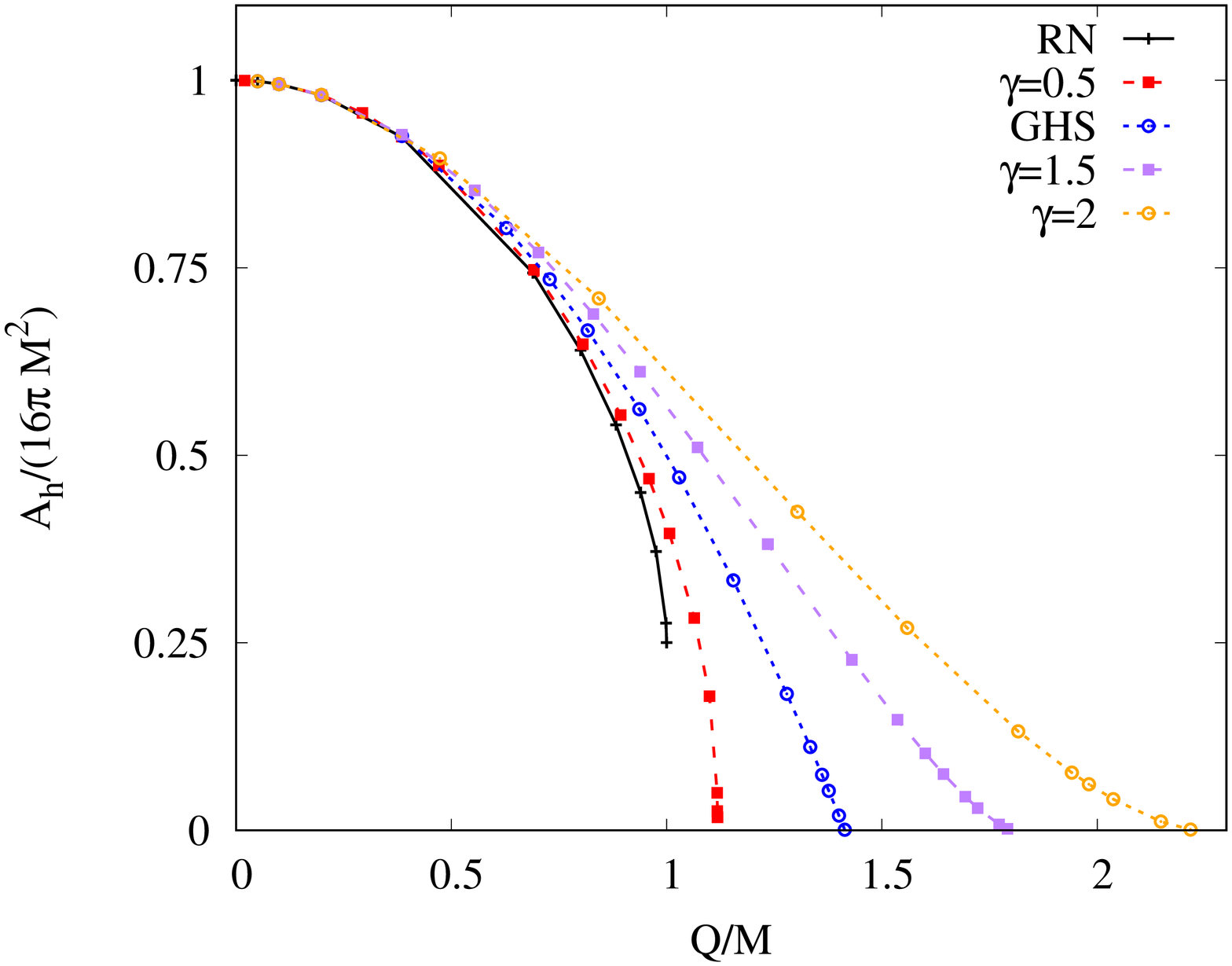}
	\caption{Static spherically symmetric black hole solutions:
(left) Dilaton charge $Q_s$ versus the electric charge $Q$, 
both scaled by the mass $M$, 
for several values of the coupling constant $\gamma$; in black for RN; 
in red, blue, purple and orange for $\gamma=0.5, 1, 1.5$ and $2$, respectively.
(right) A similar figure for the scaled horizon area $A_H$.}
	\label{fig:static}
\end{figure}

In Fig.~\ref{fig:static}(left), we show the relation 
between $Q_S/M$ and $Q/M$ for several values 
of the dilaton coupling constant $\gamma$. 
In black (solid) we show the pure EM case, 
with the RN black holes ($Q_S=0$) existing 
from $Q=0$ (Schwarzschild) to $Q=M$ (extremal).
In various colors (line styles) we show dilatonic black holes. 
All of them also emerge from the Schwarzschild solution ($Q=Q_S=0$).
But when $Q\neq0$, the scalar field becomes non-trivial
and the black holes are scalarized. 
For each $\gamma$ the solutions stop existing 
at some limiting value of the ratio $Q/M$,
which is always larger than one (overcharged solutions)
and increases with increasing $\gamma$. 
At their respective limiting value the solutions become singular: 
these purely electric black holes do not have a regular 
extremal limit in EMD theory. 
Only when allowing for a non-trivial magnetic charge 
this limit is smooth \cite{Gibbons:1987ps,Garfinkle:1990qj}.

Regarding the behavior close to the horizon, the expansion reads
\begin{eqnarray}
\label{hor_static}
f &=&f_1 \left(r-r_H\right) + O(\left(r-r_H\right)^2) \ , \nonumber \\
m &=&\frac{r_H}{2}+\frac{Q^2e^{-\gamma\phi_H}}{2r_H^2}\left(r-r_H\right) + O(\left(r-r_H\right)^2) \ , \nonumber \\
\phi_0 &=&\phi_H-\frac{2\gamma Q^2 e^{-\gamma\phi_H}}{r_H(r_H^2-Q^2e^{-\gamma\phi_H})}\left(r-r_H\right) +O(\left(r-r_H\right)^2) \ , \nonumber  \\
a_0 &=&-\Psi_H+Qe^{-\gamma\phi_H}\sqrt{\frac{f_1}{r_H(r_H^2-Q^2e^{-\gamma\phi_H})}}\left(r-r_H\right)+O(\left(r-r_H\right)^2) \ ,
\end{eqnarray}
where the black hole horizon is located at $r=r_H$.
Here the dilaton field takes the value $\phi_0(r_H)=\phi_H$,
and the electrostatic potential is $a_0(\infty)-a_0(r_H)=\Psi_H$. 
It is useful to define also
$e^{2\delta_H}=\frac{r_H^2-e^{-\gamma\phi_H}Q^2}{f_1 r_H^3}$. 
In a global solution, these near-horizon parameters 
depend in a non-trivial way on the global charges $M$ and $Q$. 
These parameters are also related to physically relevant horizon properties, 
as, for instance, the temperature $T_H$ and the area $A_H$ of the horizon
\begin{eqnarray}
T_H&=& \frac{1}{4\pi}\sqrt{\frac{f_1}{r_H^{3}}(r_H^2-Q^2e^{-\gamma\phi_H})} \ , \nonumber \\
A_H&=& 4\pi r_H^2 \ .
\end{eqnarray}

The black hole thermodynamics of static charged EMD black holes was
investigated in detail by Gibbons and Maeda \cite{Gibbons:1987ps},
who realized that the string theory value $\gamma=1$ plays a special role.
For $\gamma<1$, the temperature goes to zero
as the maximal charge is approached, analogous to the RN case.
For $\gamma>1$, the temperature diverges in this limit.
However, for $\gamma=1$, the black holes satisfy $T_H=8\pi M$, 
i.e., they approach a finite value in this limit.
In Fig.~\ref{fig:static}(right), we show the horizon area $A_H$ 
versus the electric charge (both scaled to the mass), 
for several values of the coupling constant $\gamma$. 
In the RN case (black), the maximum area is reached in the Schwarzschild case, 
and the minimum at extremality. 
In the scalarized case (colored), 
the branches of solutions end at configurations with vanishing area. 
An analysis of these solutions reveals that the limit is singular, 
where curvature invariants like the Kretschmann scalar diverge.

\section{Linear Perturbations}

In this section we present the linear perturbations of the 
previous static spherically symmetric black holes. 
Because of this symmetry of the background, 
it is convenient to study the perturbations in three different cases: 
perturbations that are purely spherical, axial (odd-parity $(-1)^{l+1}$) 
and polar (even-parity $(-1)^l$).

\subsection{Spherical perturbations}

Spherical perturbations enter the metric, the electromagnetic field 
and the scalar field. The ansatz for the metric can be written as
\begin{eqnarray}
\label{metric_l0_pert}
ds^2=-f(r) (1+\epsilon e^{-i \omega t} F_t(r)) dt^2 + \frac{1+\epsilon e^{-i \omega t} F_r(r)}{1-2m(r)/r}dr^2 +r^2(d\theta^2+\sin^2 \theta d\varphi^2) \ ,  
\end{eqnarray}
where $\epsilon$ is the control parameter for the linear expansion. 
$F_t$ and $F_r$ are radial perturbation functions, 
and $\omega=\omega_R+i\omega_I$ is the complex eigenvalue 
that parametrizes the oscillation frequency in its real part, 
and the inverse of the damping time in its imaginary part. 
Spherical perturbations for the matter can be written as
\begin{eqnarray}
\label{stab3}
A=   a_0(r) (1+\epsilon e^{-i \omega t} F_{a_0}(r)) dt \ ,~~~\phi=\phi_0(r) + \epsilon e^{-i \omega t} \phi_1(r) \ ,
\end{eqnarray} 
where $F_{a_0}$ and $\phi_1$ are the radial perturbation functions 
for the electric field and the dilaton, respectively.

Using this ansatz in the field equations 
(\ref{Eins}), (\ref{Mxw}) and (\ref{Klein}), 
and making use of the equations for the background (\ref{stateq}), 
it is possible to show that the spherical perturbations 
are described by a single Schr\"odinger-like ordinary differential equation 
for $Z=r\phi_1$ (Master equation):
\begin{eqnarray}
\frac{d^2Z}{dR^2} &=& (U_0(r) - \omega^2)Z \ , 
\end{eqnarray}
where $U_0$ is the spherical potential
\begin{eqnarray}
U_0(r)&=& 
\frac{r-2m}{2r^5e^{\gamma \phi_0 + 2\delta}}\left[\left(Q^2-r^2e^{\gamma\phi_0}\right)\left(r\partial_r\phi_0\right)^2-4\gamma Q^2 r \partial_r \phi_0 + 2Q^2(2\gamma^2-1)+4rme^{\gamma\phi_0} \right] \ ,
\end{eqnarray}
and $R$ the tortoise coordinate, for which 
\begin{eqnarray}
\partial_r R &=& \frac{1}{\sqrt{f(1-2m/r)}} \ .
\end{eqnarray}
It can actually be seen that $U_0(r)>0$ for all the black hole solutions 
we have analyzed, which immediately implies that all these solutions 
have mode stability under spherical perturbations.

In order to obtain the QNMs of the spherical perturbations, 
we have to impose the outgoing wave behavior 
as the perturbation reaches infinity. 
This means that when $r\to \infty$, we have
\begin{eqnarray}
Z =A_{\phi}^{+}e^{i\omega R} \left(1+\frac{iM}{2\omega}\frac{1}{r^2} + O(r^{-3})\right) \ .
\label{zout}
\end{eqnarray}
On the other hand, the perturbation has to be ingoing at the horizon. 
This means that when $r \to r_H$, we have
\begin{eqnarray}
Z =A_{\phi}^{-}e^{-i\omega R} \left(1 + 2r_He^{\gamma \phi_H}\frac{(i\gamma^2+e^{\delta_H}\omega r_H)Q^2-e^{\gamma\phi_H+\delta_H}\omega r_H^3}{\left(e^{\gamma\phi_H}r_H^2-Q^2\right)\left(-iQ^2+e^{\gamma\phi_H}r_H^2(i+2r_He^{\delta_H}\omega)\right)}(r-r_H) + O((r-r_H)^2)\right) \ .
\label{zin}
\end{eqnarray}
In the previous expansion, $A_\phi^{\pm}$ is an arbitrary amplitude 
for the scalar perturbation. 
The other terms in the expansion are fixed by the background solution.

\subsection{Axial perturbations}

The second type of perturbations we will study are axial, 
meaning that they transform with odd-parity under reflection 
of the angular coordinates. Because of the background symmetry, 
these perturbations enter the metric only in the following form:
\begin{eqnarray}
\label{metric_axial_pert}
ds^2&=&-f(r) dt^2 
+ \frac{1}{1-2m(r)/r}dr^2 +r^2(d\theta^2+\sin^2 \theta d\varphi^2)
\nonumber \\
 & & + 2 \epsilon h_0(r)e^{-i\omega t}\frac{\partial_{\varphi}Y_{lm}(\theta,\varphi)}{\sin{\theta}} dtd\theta +
2 \epsilon h_0(r)e^{-i\omega t}\sin{\theta}\partial_{\theta}Y_{lm}(\theta,\varphi) dtd\varphi
\nonumber \\
 & & + 2 \epsilon h_1(r)e^{-i\omega t}\frac{\partial_{\varphi}Y_{lm}(\theta,\varphi)}{\sin{\theta}} drd\theta 
+
2 \epsilon h_1(r)e^{-i\omega t}\sin{\theta}\partial_{\theta}Y_{lm}(\theta,\varphi) drd\varphi \ ,
\end{eqnarray}
where now $h_0$ and $h_1$ are the radial perturbation functions, 
and $Y_{lm}$ are the standard spherical harmonics. 
The axial perturbations also enter the electromagnetic field
\begin{eqnarray}
\label{em_axial_pert}
A= a_0(r) dt -\epsilon W_2(r)e^{-i\omega t} \frac{\partial_{\varphi}Y_{lm}(\theta,\varphi)}{\sin{\theta}}d\theta
+ \epsilon W_2(r)e^{-i\omega t} \sin{\theta}{\partial_{\theta}Y_{lm}(\theta,\varphi)} d\varphi \ , 
\end{eqnarray}
where the perturbation function $W_2$ is introduced.

Using this ansatz in the field equations 
(\ref{Eins}), (\ref{Mxw}) and (\ref{Klein}), 
a set of coupled differential equations is obtained,
exhibited in the Appendix.
It consists of
two first order equations for $h_0$ and $h_1$,
and a second order equation for $W_2$.
We can write this system in the form
\begin{eqnarray}
\partial_r \Psi_{A} = M_{A} \Psi_{A} \ ,
\end{eqnarray}
where 
\begin{eqnarray}
\Psi_A = \left[ \begin{array}{c}
h_0
\\ 
h_1
\\ 
W_2
\\ 
\partial_r W_2
\end{array} 
\right] \ ,
\end{eqnarray}
and $M_{A}$ is a complicated $4\times 4$ matrix
that depends on the background metric functions and fields,
the $l$ number, and the complex eigenvalue of the mode $\omega$.

Space-time perturbations are parametrized by $\{h_0, h_1\}$,
while electromagnetic perturbations are parametrized
by $\{W_2, \partial_r W_2\}$.
Apart from the background functions,
the only explicit coupling between the space-time perturbations
and the electromagnetic perturbations appears in equation (\ref{eqh0})
in the first term, which is essentially proportional to $Q$.

In the Schwarzschild case, the system decouples into two sets
(composed of two first order differential equations).
One set is equivalent to the Regge-Wheeler equation
for axial space-time perturbations,
while the other is equivalent to the perturbation equation
for purely axial electromagnetic perturbations.
In general, when the black hole is charged, the system is fully coupled.

In order to obtain the QNMs of the axial perturbations,
we again need to impose the outgoing wave behavior at infinity
and ingoing wave behavior close to the horizon,
as specified in the Appendix.

\subsection{Polar perturbations}

The third type of perturbations we will study are polar perturbations, 
meaning they transform evenly under reflection of the angular coordinates. 
These perturbations enter the metric, the electromagnetic field 
and the scalar field.
(Note that the spherical perturbations we have described already 
would correspond to the following $l=0$ case, 
but with a slightly different gauge, 
hence it is convenient to differentiate between them.)

The ansatz for the perturbations of the metric can be written as
\begin{eqnarray}
\label{metric_polar_pert}
ds^2&=&-f(r) (1+\epsilon e^{-i \omega t} N(r) Y_{lm}(\theta,\varphi)) dt^2 
- 2\epsilon e^{-i\omega t} H_1(r) Y_{lm}(\theta,\varphi) dtdr
+ \frac{1-\epsilon e^{-i \omega t} L(r)Y_{lm}(\theta,\varphi)}{1-2m(r)/r}dr^2
\nonumber \\
& &+(r^2-2\epsilon e^{-i \omega t} T(r)Y_{lm}(\theta,\varphi))(d\theta^2+\sin^2 \theta d\varphi^2) \ ,
\end{eqnarray}
where we have introduced the perturbation functions $N$, $H_1$, $L$ and $T$. 
For the electromagnetic field we have
\begin{eqnarray}
\label{A_polar}
A&=& ( a_0(r)+\epsilon e^{-i \omega t} a_{1}(r) Y_{lm}(\theta,\varphi)) dt 
+ \epsilon W_1(r) e^{-i \omega t} Y_{lm}(\theta,\varphi) dr
 \nonumber \\
& &+ \epsilon V_1(r) e^{-i \omega t} \partial_{\theta}Y_{lm}(\theta,\varphi) d\theta + \epsilon V_1(r) e^{-i \omega t} \partial_{\varphi}Y_{lm}(\theta,\varphi) d\varphi \ ,
\end{eqnarray} 
with the perturbation functions $a_1$, $W_1$ and $V_1$. 
Finally, for the scalar field we have
\begin{eqnarray}
\label{s_polar}
\phi&=&\phi_0(r) + \epsilon e^{-i \omega t} \phi_1(r) Y_{lm}(\theta,\varphi) \ ,
\end{eqnarray} 
with the scalar perturbation function $\phi_1$.

As before, we can insert this ansatz into the field equations 
(\ref{Eins}), (\ref{Mxw}) and (\ref{Klein}), 
resulting in a set of coupled differential equations,
presented in the Appendix.
By introducing several field re-definitions $F_0$, $F_1$ and $F_2$
in terms of the above perturbation functions $W_1$, $V_1$ and $a_1$,
see Appendix (\ref{redef}),
the system of equations can be simplified. 
It is then possible to show 
that the minimal set of differential equations (Master equations) 
can be written in vectorial form as
\begin{eqnarray}
\partial_r \Psi_P = M_P \Psi_P \ ,
\end{eqnarray} 
with the vector
\begin{eqnarray}
\Psi_P = \left[ \begin{array}{c}
H_1
\\ 
T
\\ 
F_0
\\ 
F_1
\\
\phi_1
\\
\partial_r \phi_1
\end{array} 
\right] \ ,
\end{eqnarray}
and $M_P$ being a complicated $6 \times 6$ matrix, 
whose components depend on the background metric functions and fields, 
the $l$ number and the mode $\omega$.

Space-time perturbations are parameterized by $\{H_1,T\}$,
while electromagnetic perturbations by $\{F_0,F_1\}$,
and scalar perturbations by $\{\phi_1,\partial_r \phi_1\}$.
In the Schwarzschild case, the system decouples into three sets of equations
(composed of two coupled first order equations each).
One set is equivalent to the Zerilli equation for
polar space-time perturbations, while the other two are equivalent
to the perturbation equation for purely polar electromagnetic perturbations
and the perturbation equation for a minimally coupled scalar field
in the Schwarzschild background, respectively.

When the black hole is charged (RN),
the equations for the space-time perturbations couple with the equations
for the electromagnetic perturbations,
but the equations for the scalar perturbations are decoupled.
In the general case we are considering here, where the black hole is
electrically charged and also carries a nontrivial scalar field,
all equations are coupled to each other.

Again, in order to obtain the QNMs, we need to impose 
the outgoing wave behavior at infinity and
ingoing wave behavior close to the horizon,
as explicitly shown in the Appendix.

\section{Numerical results}

\subsection{Overview of the method and results}

In order to obtain the QNMs for dilatonic RN black holes 
for general dilaton coupling constant $\gamma$ 
and any associated allowed values of the mass $M$ and electric charge $Q$,
we implement the method described in the following. 

First, we generate numerically the background solutions with high precision. 
For this, we solve numerically the equations (\ref{stateq}),
imposing the boundary conditions resulting from expansions (\ref{inf_static}) 
and (\ref{hor_static}), employing the ordinary differential equation
solver COLSYS \cite{Ascher:1979iha}. 

For the calculation of the QNMs, 
we follow a procedure similar to the one previously 
used in other cases \cite{Blazquez-Salcedo:2016enn,Blazquez-Salcedo:2017txk}. 
The space-time is divided in two regions: 
region I, from $r=r_H+\epsilon_H$ to $r=r_J$, 
and region II from $r=r_J$ to $r=r_\infty>r_J$. 
In region I, we parametrize the ingoing wave behavior 
(for the radial perturbations given by expression (\ref{zin}), 
for axial by (\ref{axin}) and for polar by (\ref{polin})). 
Similarly, in region II, we parametrize the outgoing wave behavior 
(which now for the radial perturbations is given by expression (\ref{zout}), 
for axial by (\ref{axout}) and for polar by (\ref{polout})). 
We generate numerically sets of linearly independent solutions 
and match them at $r_J$. 
QNMs with eigenvalue $\omega$ are found
when the matching of the functions and their derivatives is continuous.

In practice, our numerical implementation allows us to connect 
the dilatonic solutions with the pure RN black hole solutions continuously 
(for example, by generating families of charged solutions 
by slowly increasing the coupling constant from $\gamma=0$ 
to any arbitrary value of $\gamma$).
This is convenient because we can continuously track the QNMs, 
and connect them to all the known spectra 
(i.e., those of Schwarzschild, RN and GHS black holes). 
This allows us to cross-check all numerical calculations of the QNMs.

Our results for the QNMs of the RN black holes 
reproduce the results in \cite{Leaver:1990zz,Richartz:2014jla}. 
For the GHS black hole ($\gamma=1$), our results reproduce the QNMs 
calculated in \cite{Ferrari:2000ep}. All these modes are stable.

In the following, we will comment on our results 
for the QNM spectrum of the dilatonic RN black holes 
with arbitrary coupling $\gamma$. 
In particular, the modes can be categorized into three different families: 
\begin{itemize}
\item[i.]
We call modes that can be connected with purely gravitational perturbations 
of the Schwarzschild solution gravitational-led (grav-led) modes. 
Typically, these perturbations are led by space-time oscillations 
with the dominant amplitude $A_g^{\pm}$. 
\item[ii.]
We call modes that can be connected with purely electromagnetic perturbations 
of the Schwarzschild solution electromagnetic-led (EM-led) modes. 
In this case, the perturbations are led by oscillations 
of the electromagnetic field with the dominant amplitude $A_F^{\pm}$.
\item[iii.]
We call modes that can be connected with purely scalar perturbations 
of the Schwarzschild solution scalar-led modes, corresponding to
a minimally coupled scalar field in this background. 
Here the perturbations are led by oscillations of the scalar field
with the dominant amplitude $A_\phi^{\pm}$.
\end{itemize}

Of course, in the general case, where the background 
is electrically charged and has dilatonic hair, 
all the perturbations are coupled with each other. 
The stronger the coupling, i.e. the larger $Q$ and $Q_S$, 
the stronger is the coupling of the perturbations.
Nonetheless, grav-led modes only appear for $l\ge 2$, 
since they correspond to tensor perturbations of the metric. 
EM-led modes appear for $l \ge 1$, 
since they correspond to vector perturbations. 
Scalar-led modes appear for $l \ge 0$, 
since obviously, they correspond to scalar perturbations. 

On top of this distinction referring to the physical origin of each mode, 
we have the two decoupled channels of perturbations 
with (in principle) their own modes: axial and polar. 
However, it is well-known that in EM theory, both spectra coincide, 
which is called isospectrality of the QNMs of the RN black holes. 
As demonstrated in the following for general dilaton coupling $\gamma$
and general electric charge $Q$ (below the respective maximal value),
in the presence of a non-trivial dilaton, isospectrality is broken
(as shown in \cite{Ferrari:2000ep} for the case $\gamma=1$
and for general $\gamma$ but small values of the charge in \cite{Brito:2018hjh}).
In particular, we will now discuss our results for every $l$ number separately,
devoting a subsection to $l=0,1$ and 2 each, and selecting for the
dilaton coupling constant always the values $\gamma=0, 0.5, 1, 1.5$.

\subsection{Spectrum of $l=0$ perturbations}

\begin{figure}[t]
	\centering
	\includegraphics[width=0.35\linewidth,angle=-90]{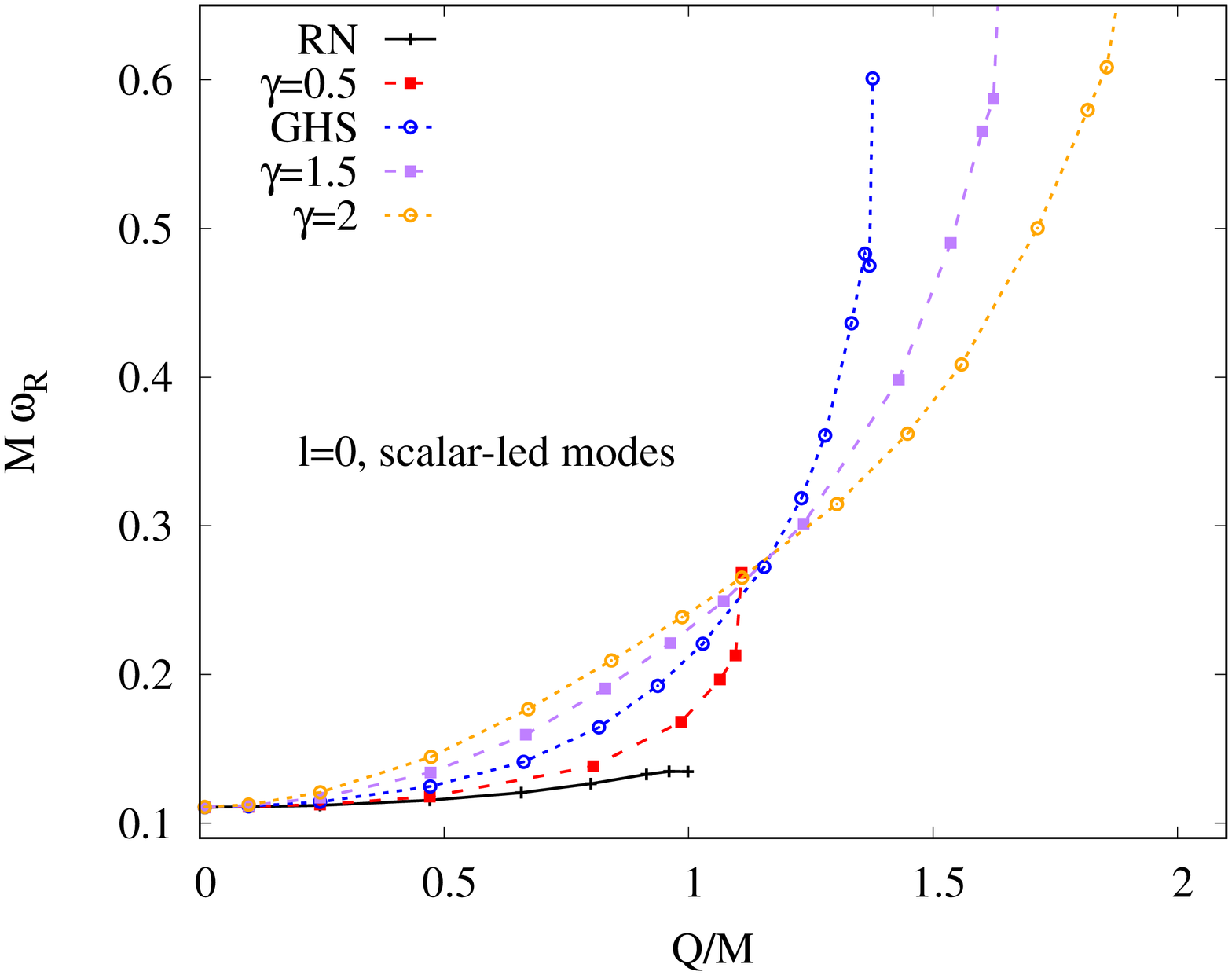}
		\includegraphics[width=0.35\linewidth,angle=-90]{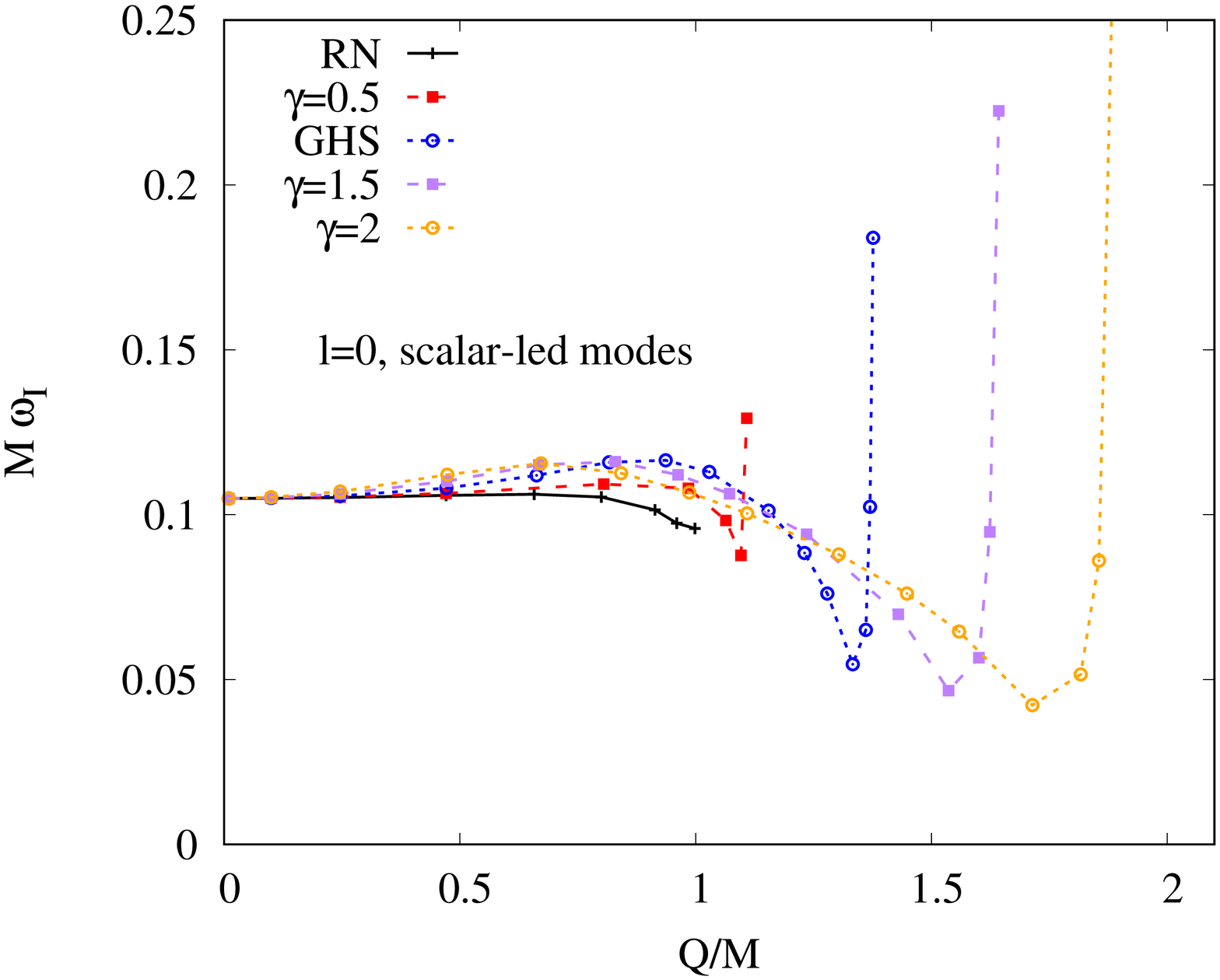}
	\caption{Scalar-led modes for radial $l=0$ perturbations.}
	\label{fig:l0_scalar_R}
\end{figure}

The $l=0$ perturbations possess a single family of scalar-led modes.
In Fig.~\ref{fig:l0_scalar_R}(left), we show the real part of the frequency $\omega_R$ (scaled by the mass) 
as a function of the electric charge $Q$  (also scaled to the mass). 
Figure \ref{fig:l0_scalar_R}(right) shows the imaginary part $\omega_I$ (scaled by the mass) of the modes. 
In black we show the RN modes, and in colors several values of the dilaton coupling constant $\gamma$: 
in red for $\gamma=0.5$, in blue for the GHS solution with $\gamma=1$, in purple for $\gamma=1.5$ 
and in orange for $\gamma=2$. 

In the RN case, the real and imaginary parts of the modes do not deviate much from the Schwarzschild 
mode when the charge $Q/M$ is increased, even up  to  the extremal limit $Q=M$. 
The modes of dilatonic black holes deviate much more from the Schwarzschild modes. 
This is to be expected, since these black holes have a non-trivial spherically symmetric dilaton background field. 
The higher the coupling constant $\gamma$, the larger is typically the deviation from the GR spectrum, 
in particular close to the respective critical solution with maximal $Q/M$. 

The figure shows that the real part $\omega_R$ increases monotonically with increasing $Q/M$ and 
increasing $\gamma$. The imaginary part $\omega_I$, however, is not monotonic.
As $Q/M$ increases, at first $\omega_I$ increases as well, reaches a maximum, then decreases to a minimum,
which for the larger values of $\gamma$ has roughly a value of $\omega_I\simeq 0.05$, from where
it rises steeply as the maximal $Q/M$ is approached. 
Thus, for large scalarization of the black holes, the configurations can possess radial modes 
with damping times twice as large as in GR, but also much shorter damping times.

\subsection{Spectrum of $l=1$ perturbations}

\begin{figure}[h!]
	\centering
	\includegraphics[width=0.35\linewidth,angle=-90]{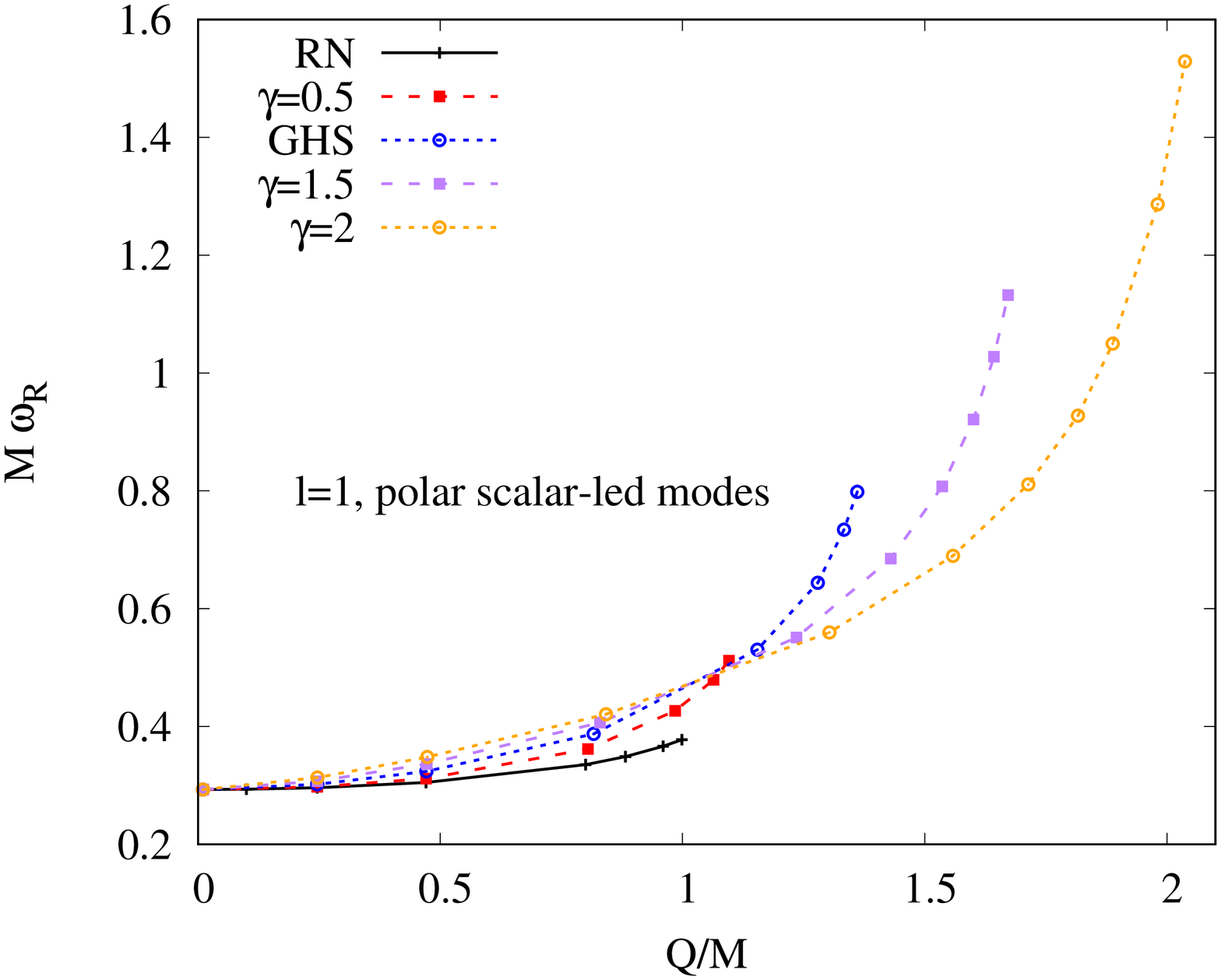}
	\includegraphics[width=0.35\linewidth,angle=-90]{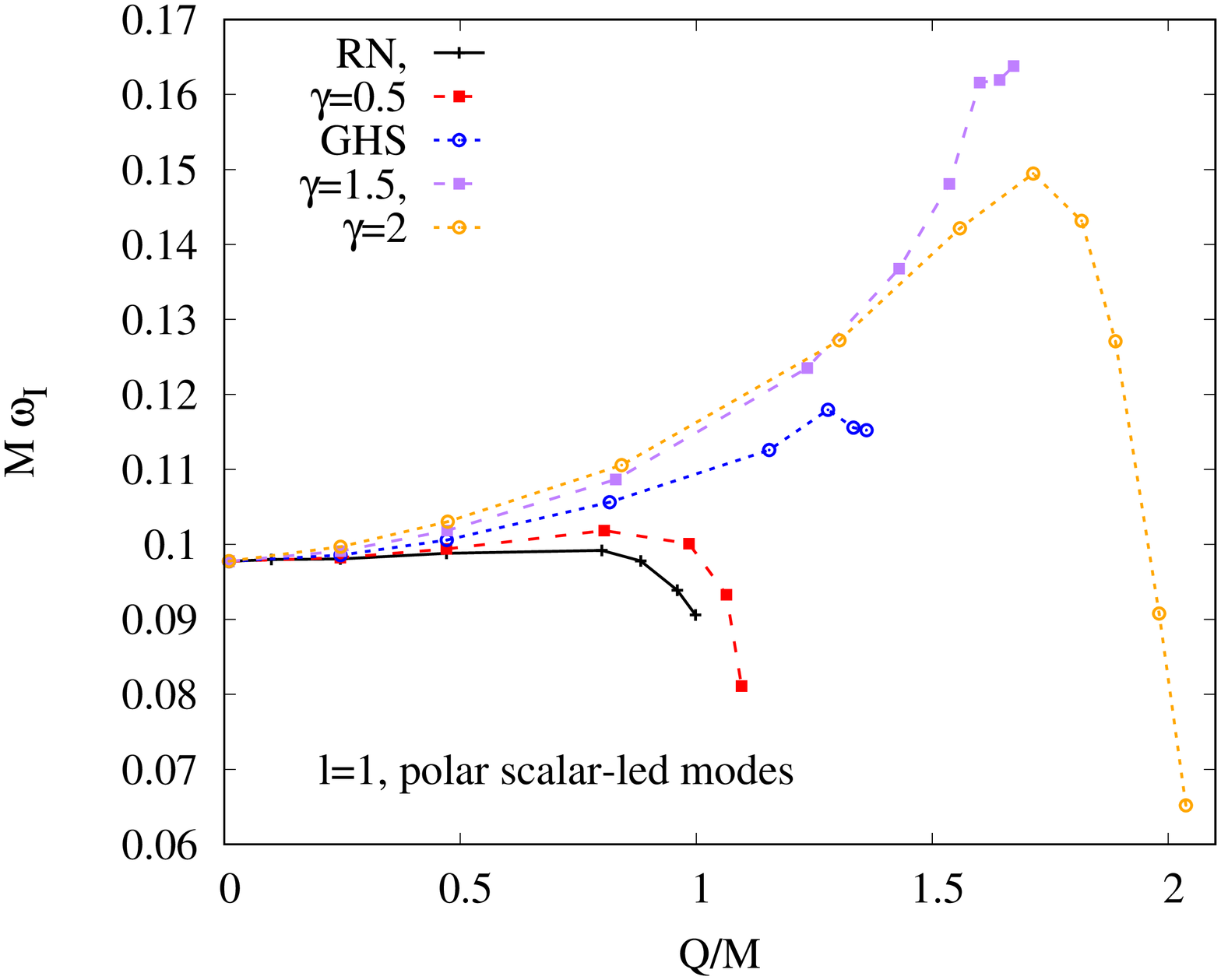}
	\caption{Scalar-led modes for polar $l=1$ perturbations.}
	\label{fig:l1_polar_scalar_R}
	\includegraphics[width=0.35\linewidth,angle=-90]{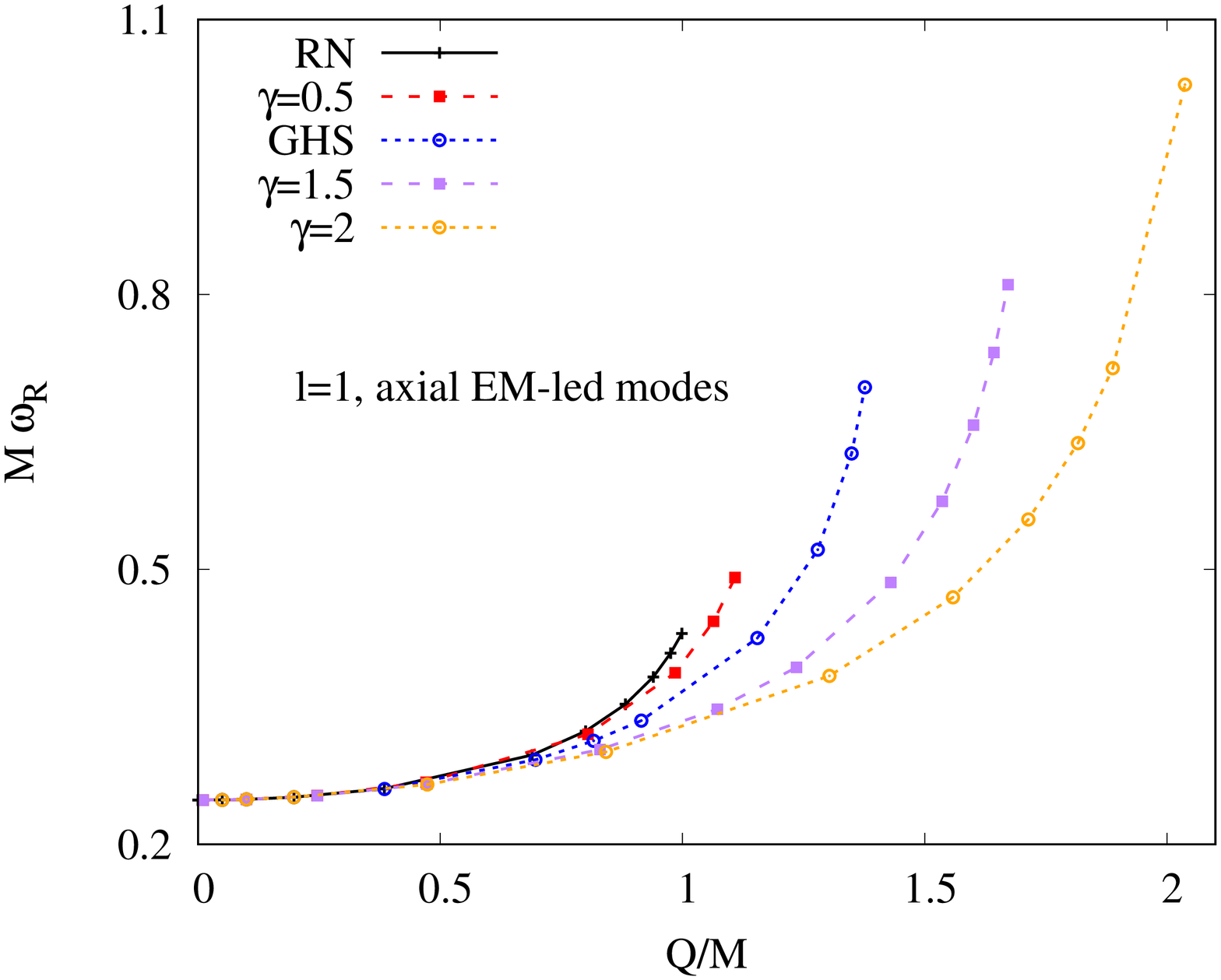}
	\includegraphics[width=0.35\linewidth,angle=-90]{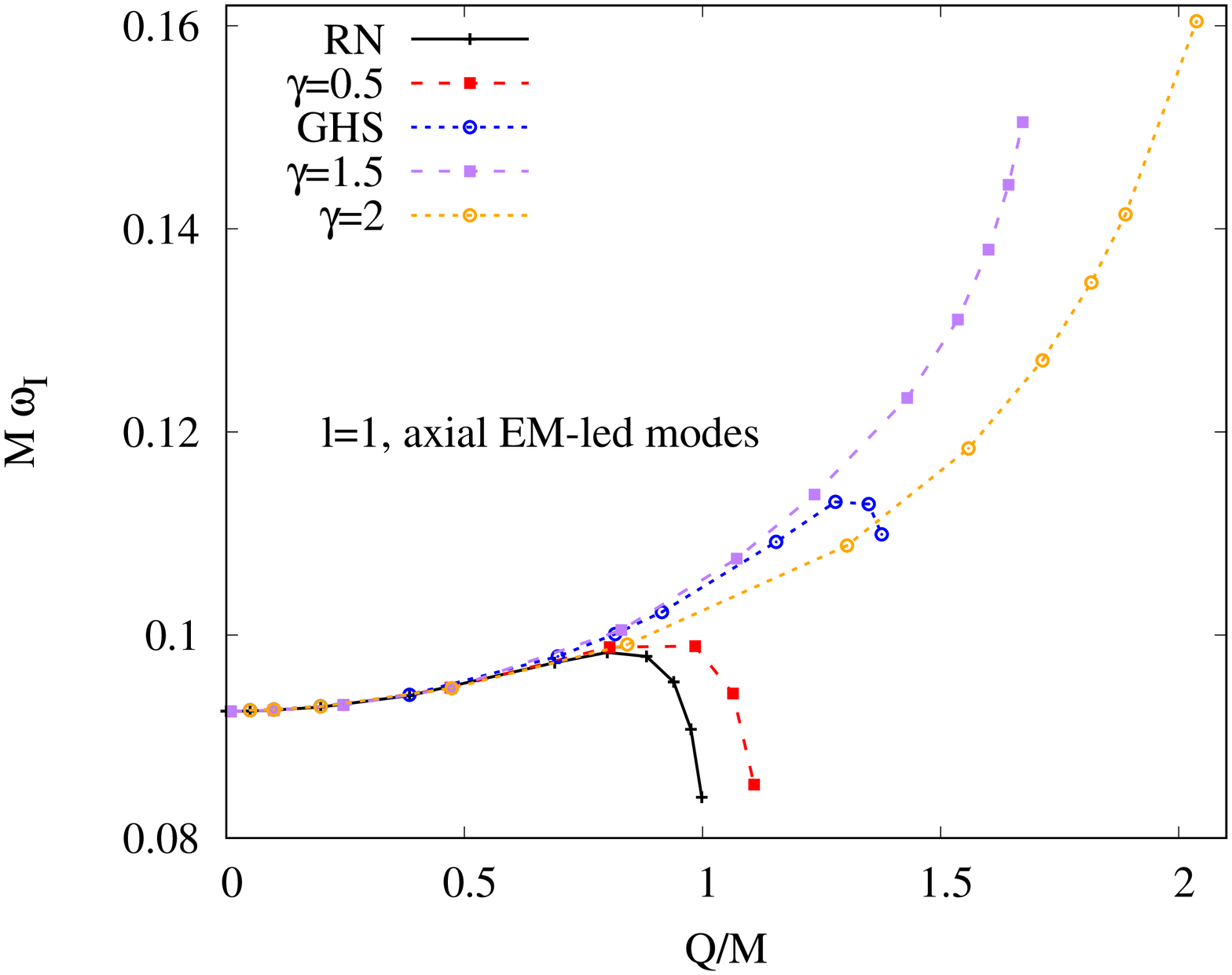}
	\includegraphics[width=0.35\linewidth,angle=-90]{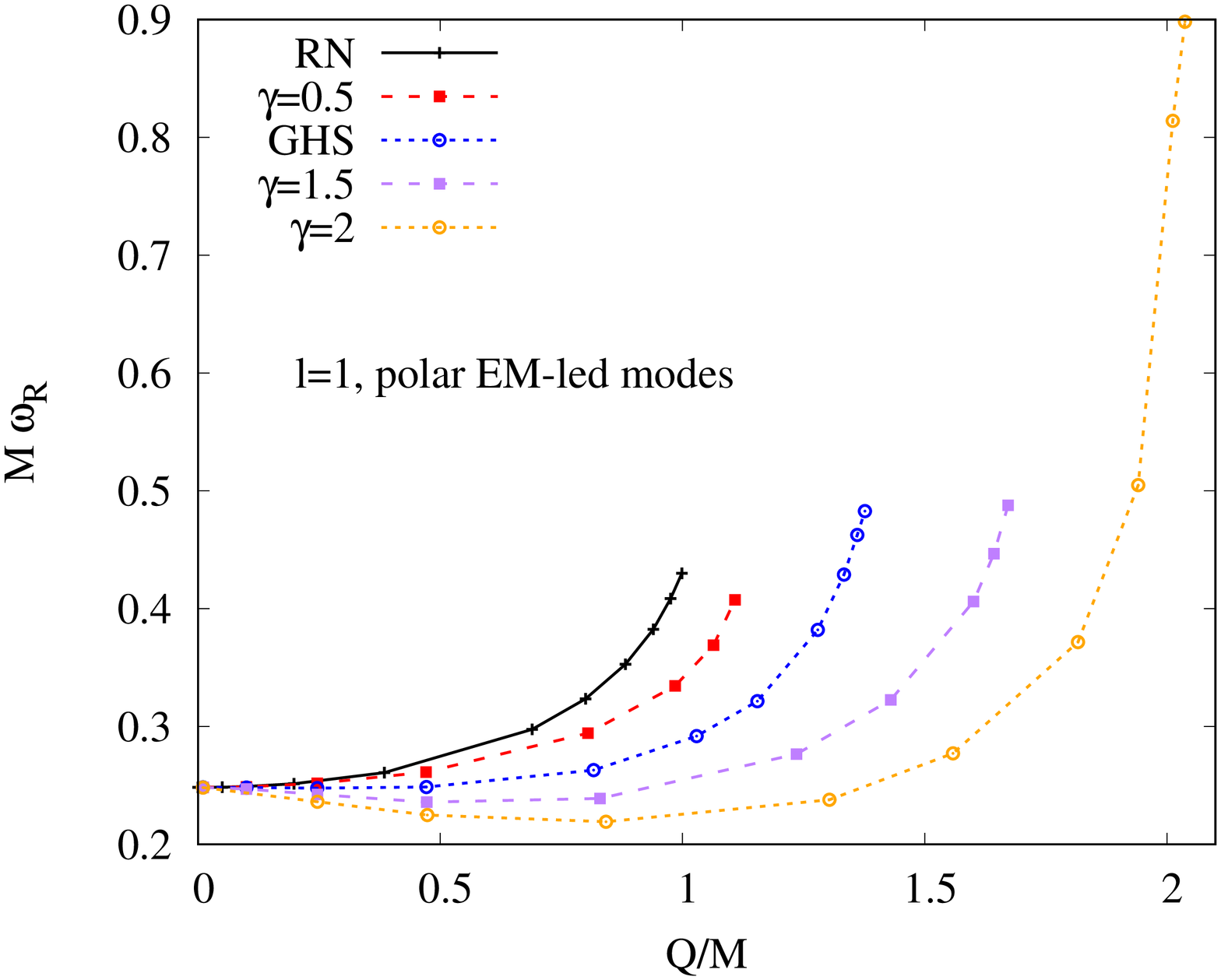}
	\includegraphics[width=0.35\linewidth,angle=-90]{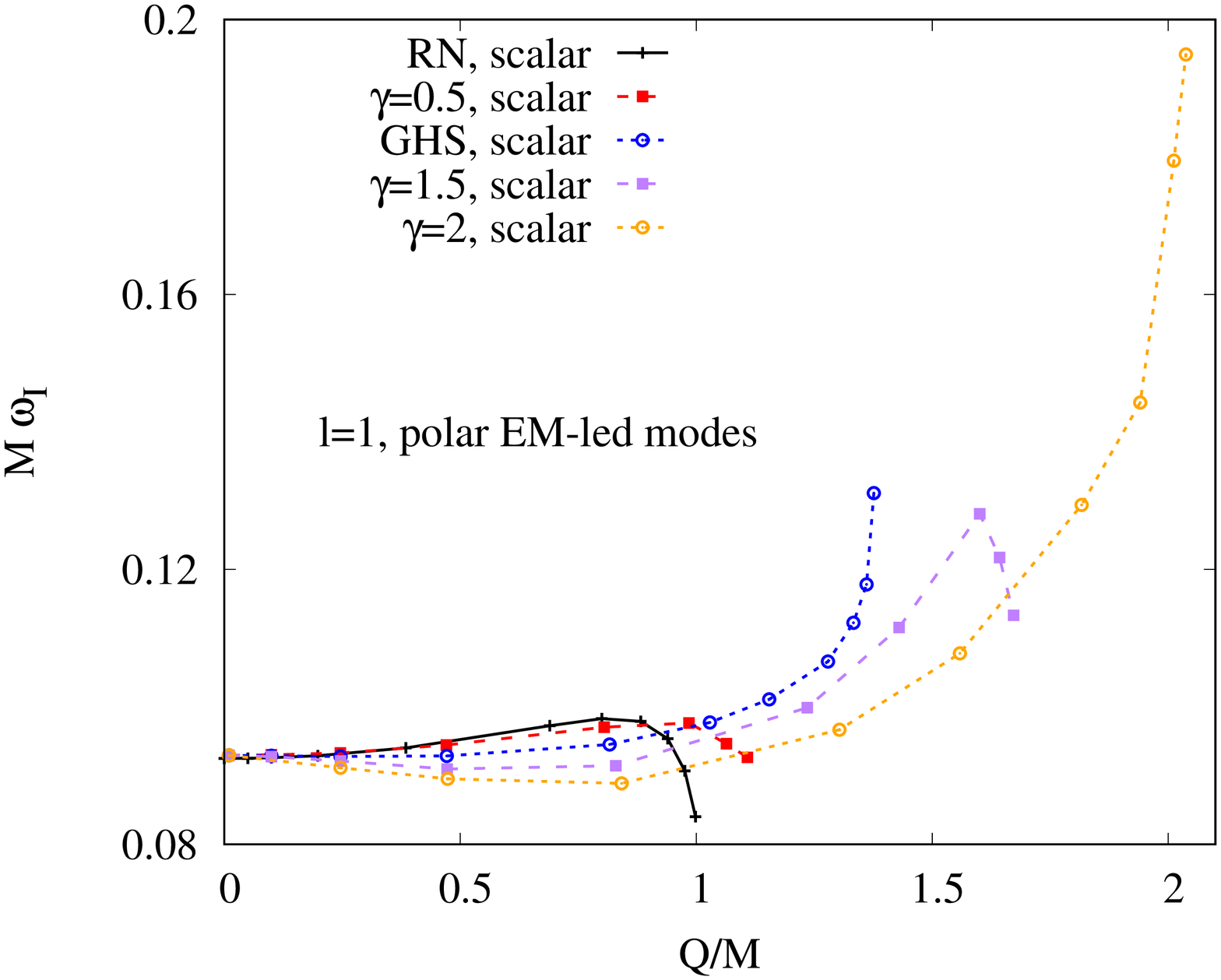}
	\caption{EM-led modes for $l=1$ perturbations:
axial (upper) and polar (lower) modes.}
	\label{fig:l1_axial_polar_EM_R}
\end{figure}

\begin{figure}[t!]
	\centering
	\includegraphics[width=0.35\linewidth,angle=-90]{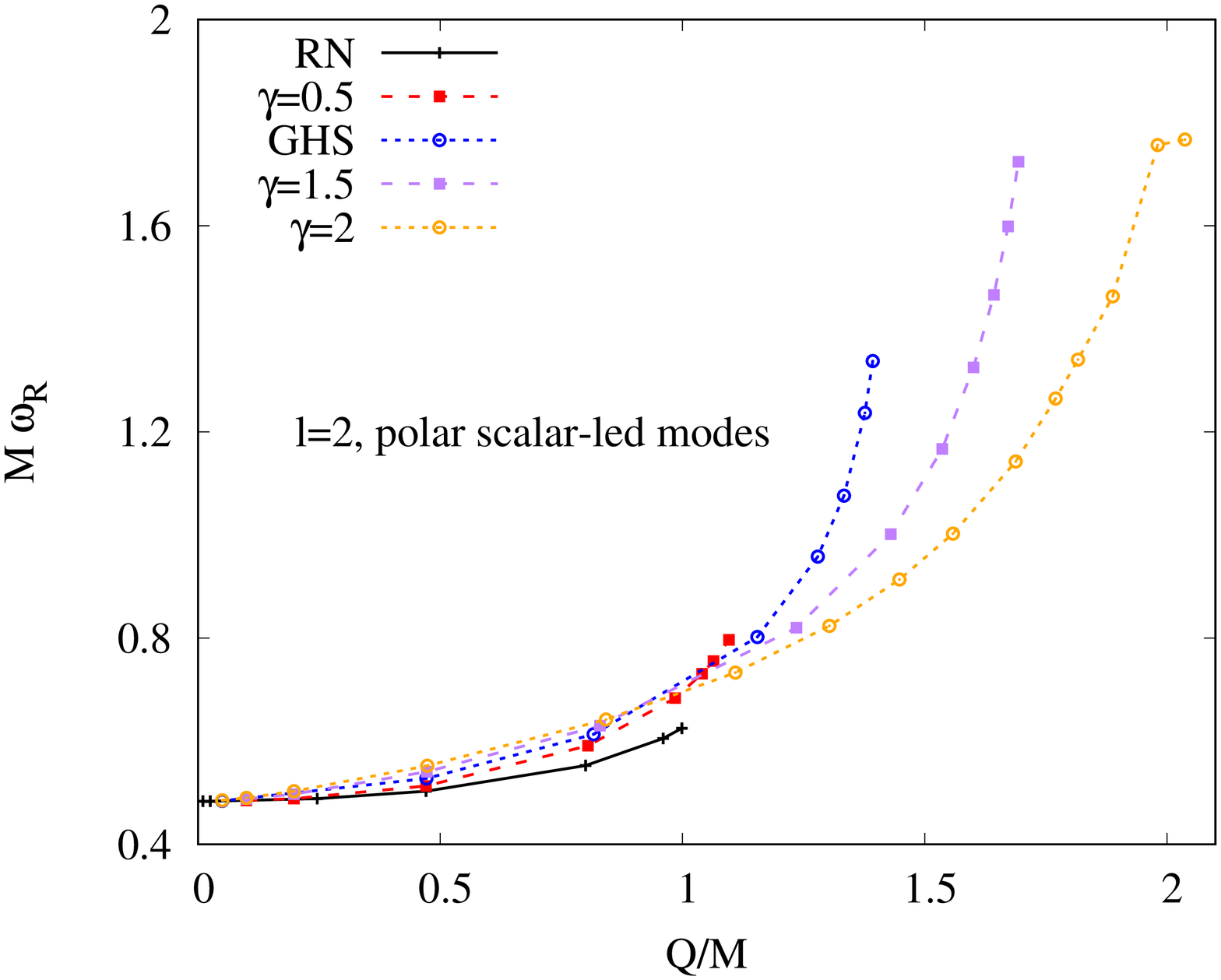}
	\includegraphics[width=0.35\linewidth,angle=-90]{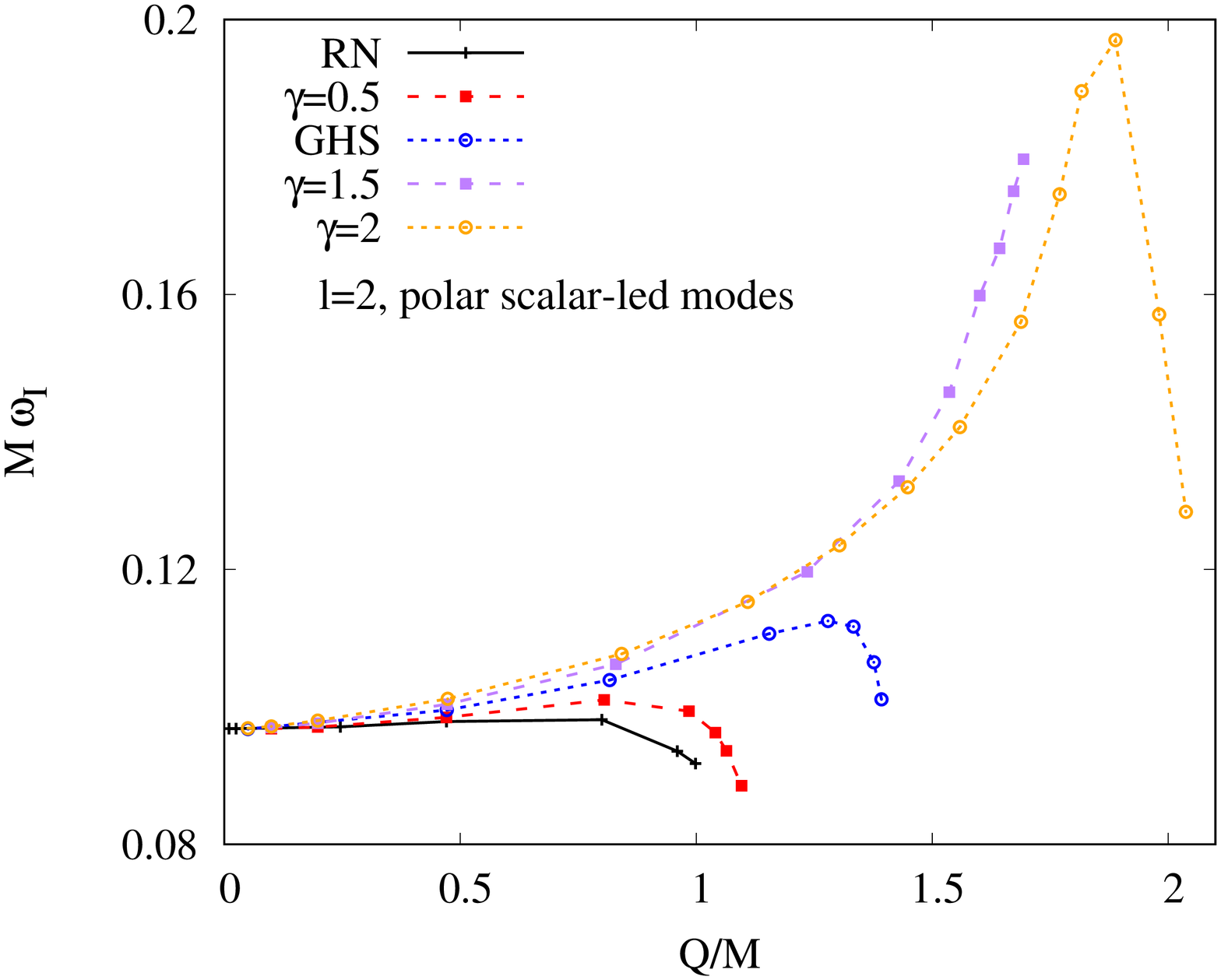}
	\caption{Scalar-led modes for polar $l=2$ perturbations.}
	\label{fig:l2_polar_scalar_R}
	\includegraphics[width=0.35\linewidth,angle=-90]{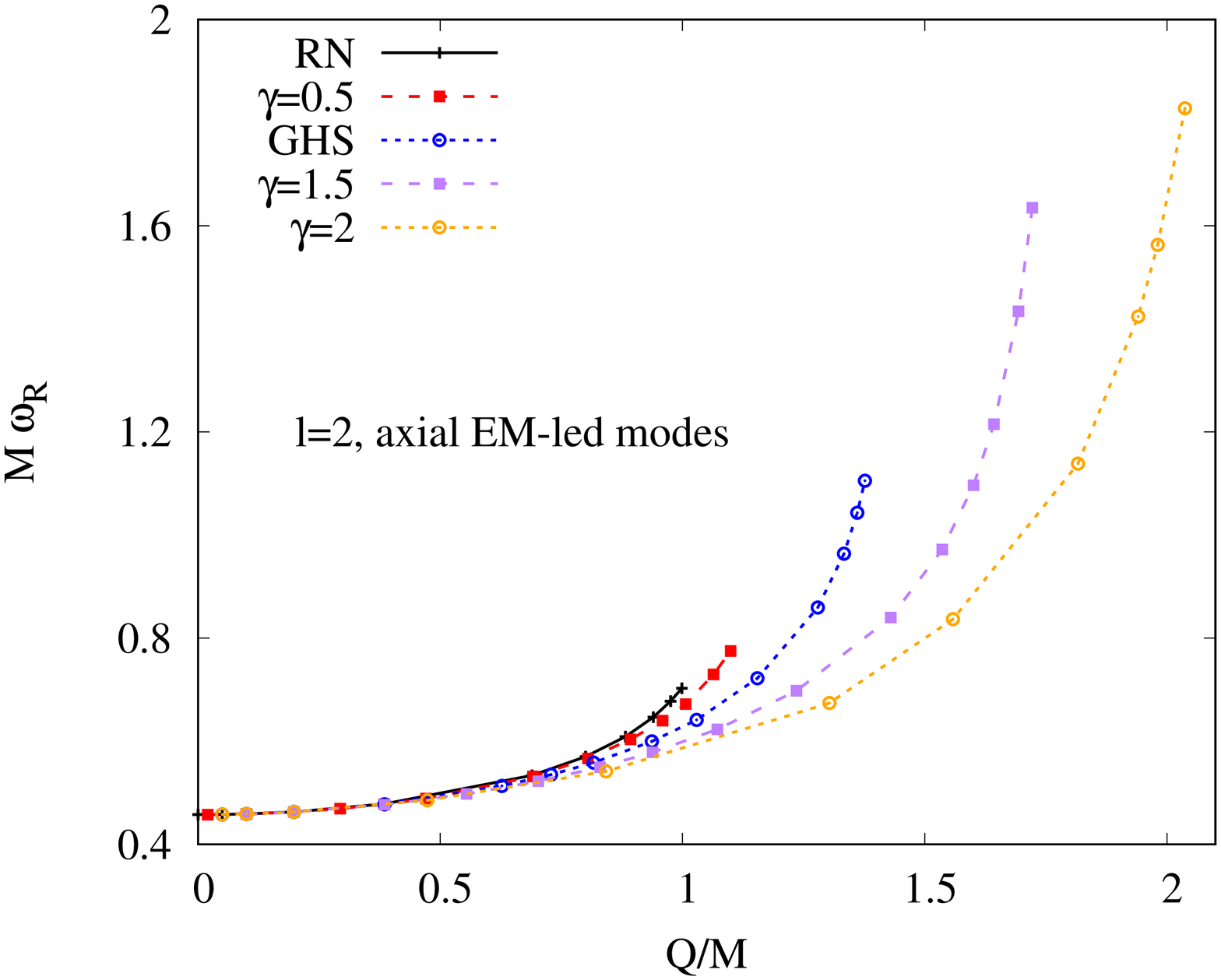}
	\includegraphics[width=0.35\linewidth,angle=-90]{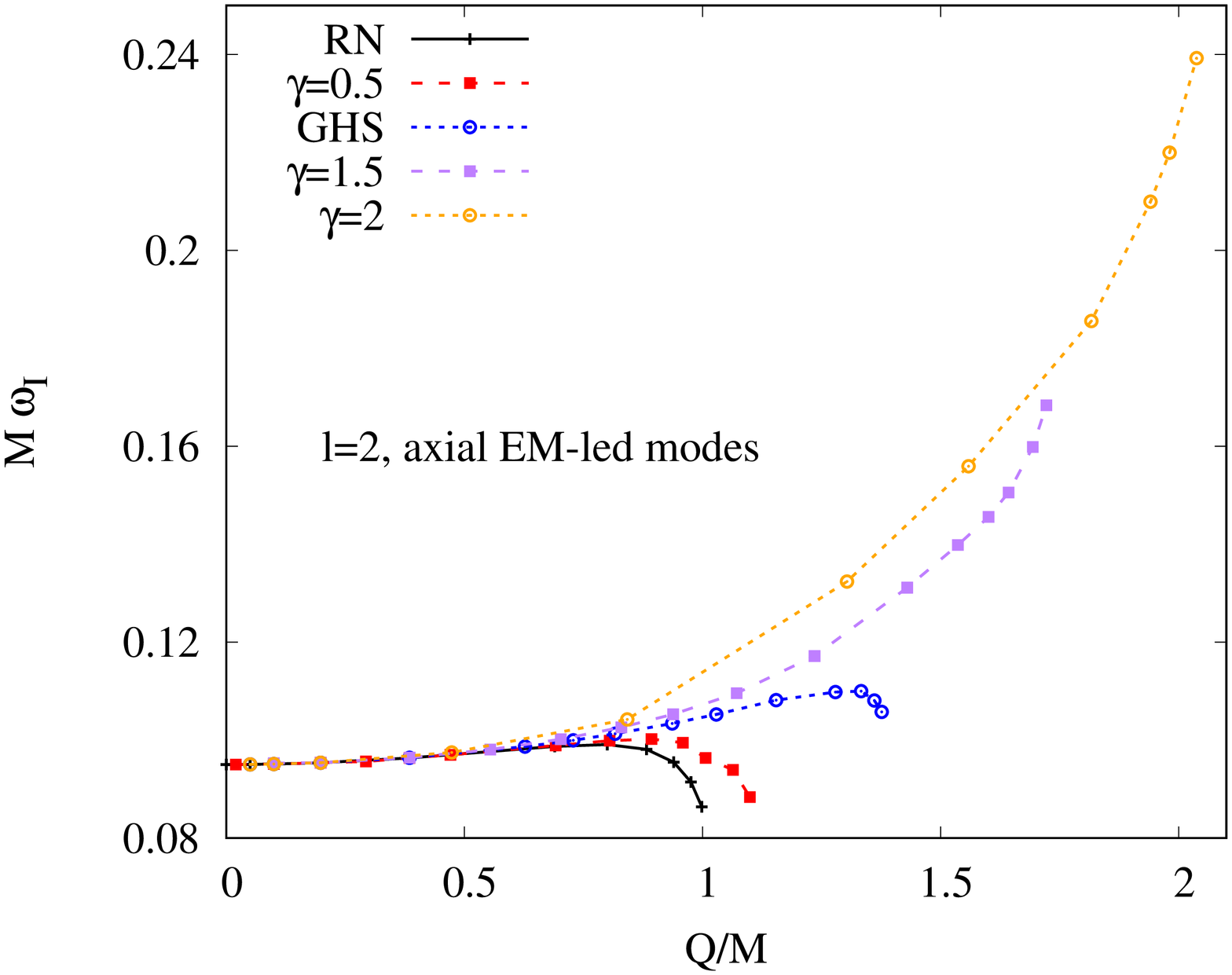}
	\includegraphics[width=0.35\linewidth,angle=-90]{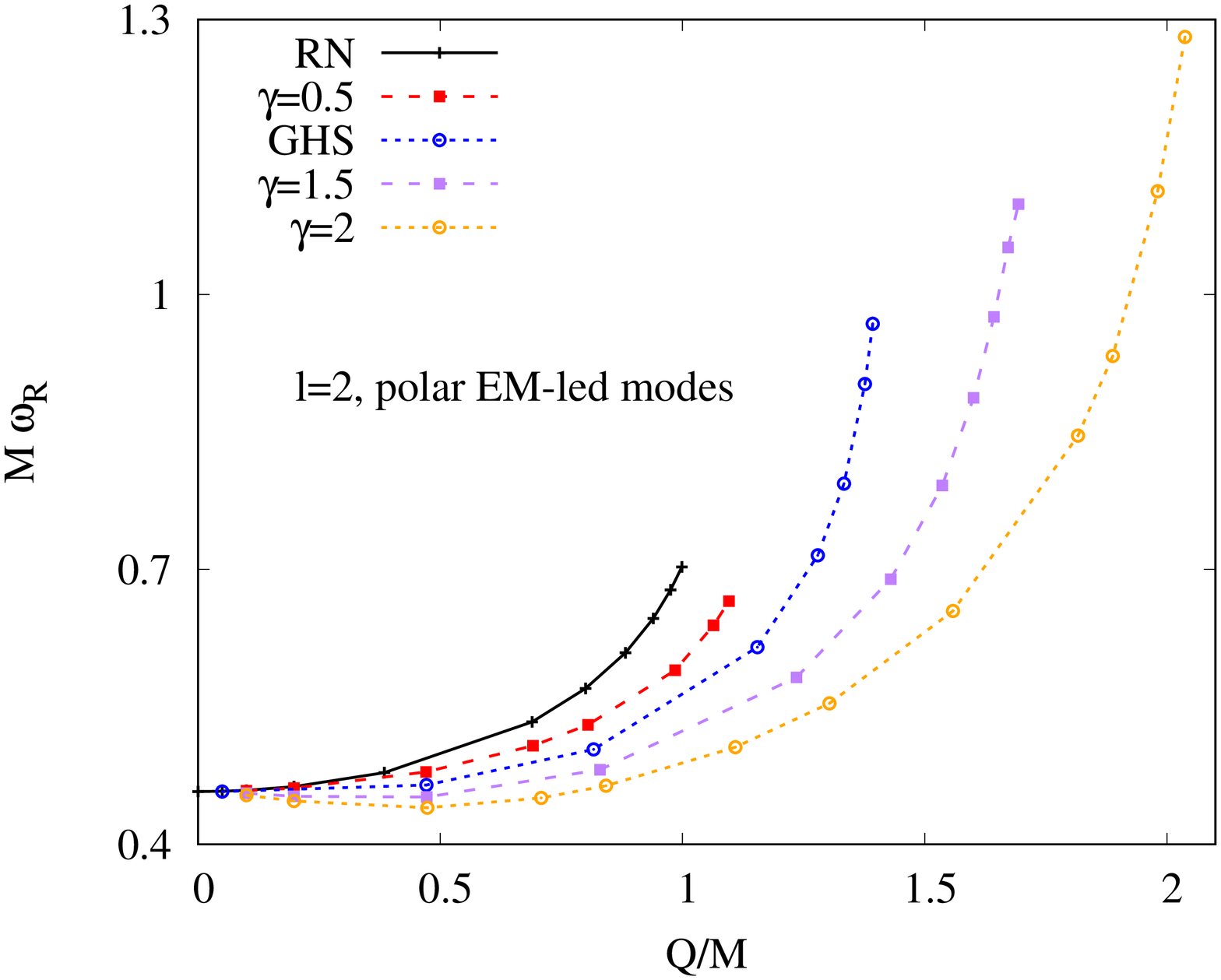}
	\includegraphics[width=0.35\linewidth,angle=-90]{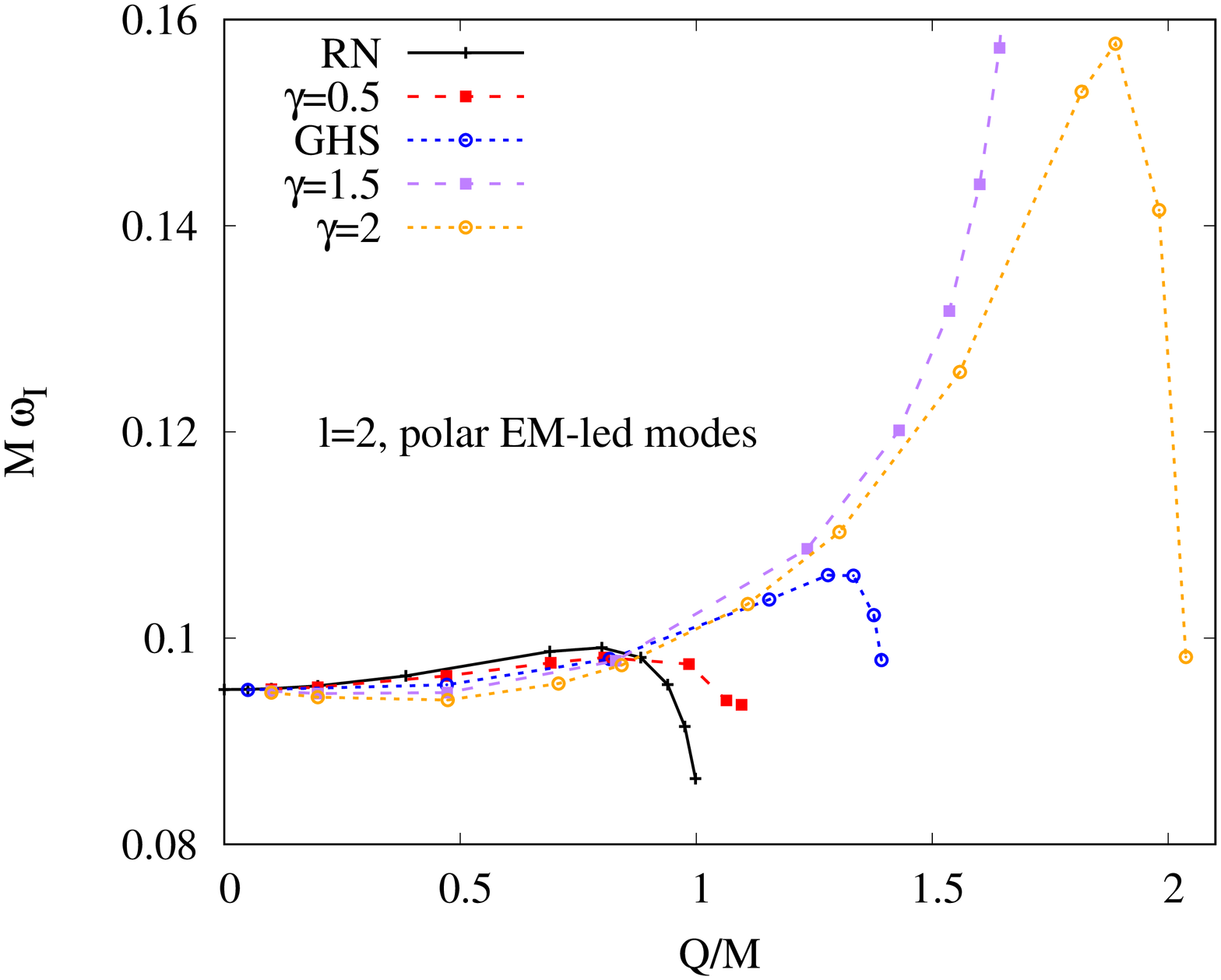}
	\caption{EM-led modes for $l=2$ perturbations:
axial (upper) and polar (lower) modes.}
	\label{fig:l2_axial_polar_EM_R}
\end{figure}
\begin{figure}[h!]
	\centering
	\includegraphics[width=0.35\linewidth,angle=-90]{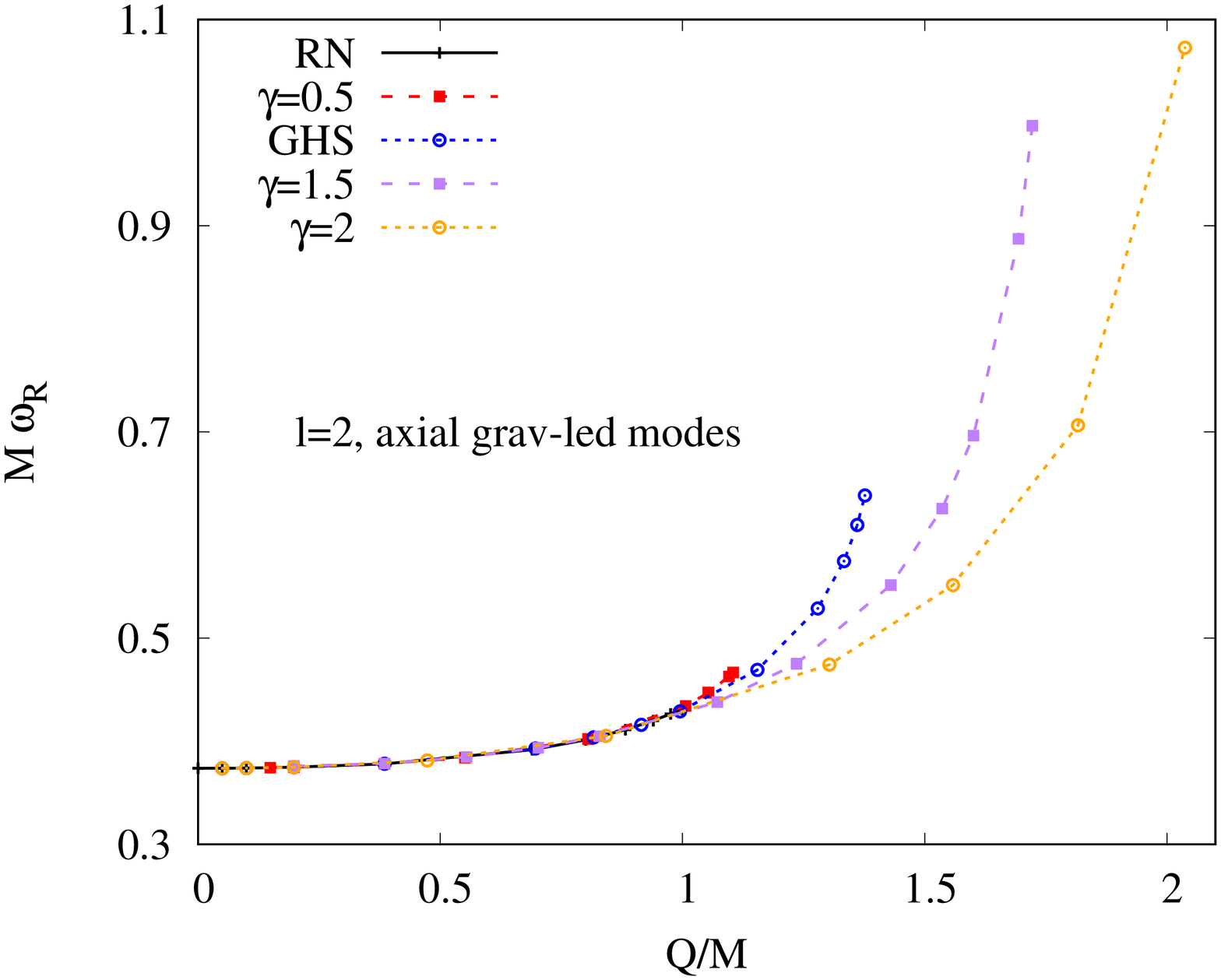}
	\includegraphics[width=0.35\linewidth,angle=-90]{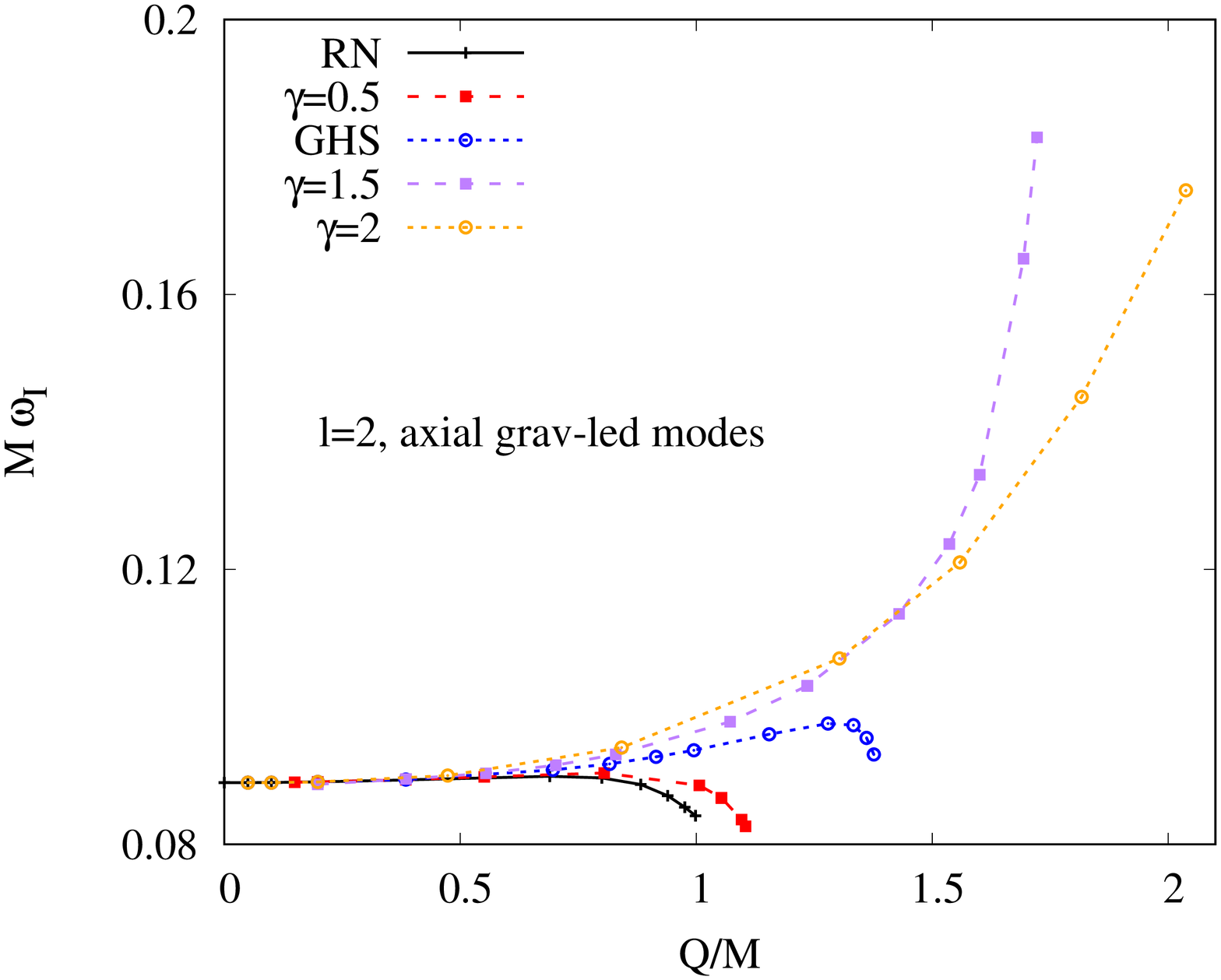}
	\includegraphics[width=0.35\linewidth,angle=-90]{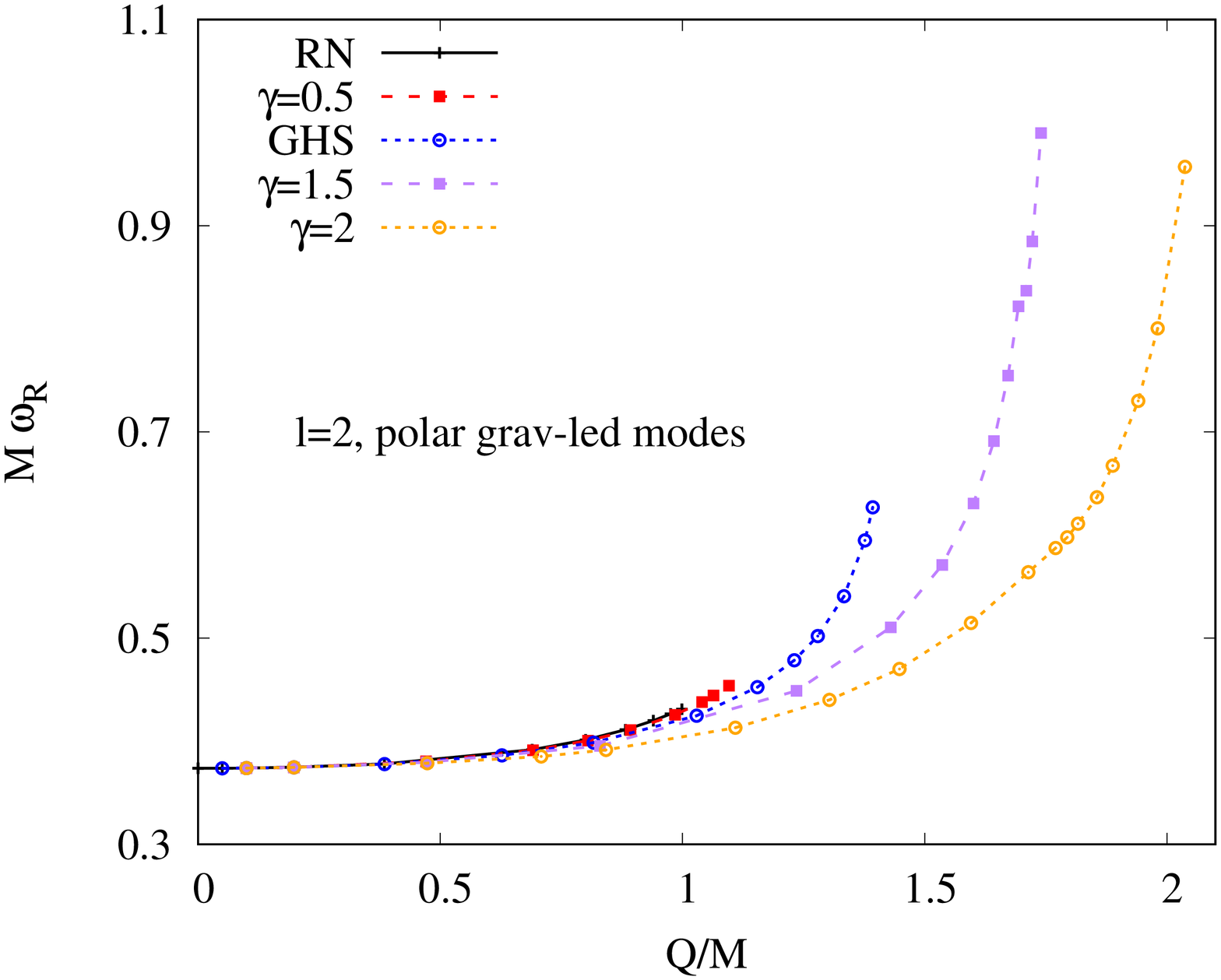}
	\includegraphics[width=0.35\linewidth,angle=-90]{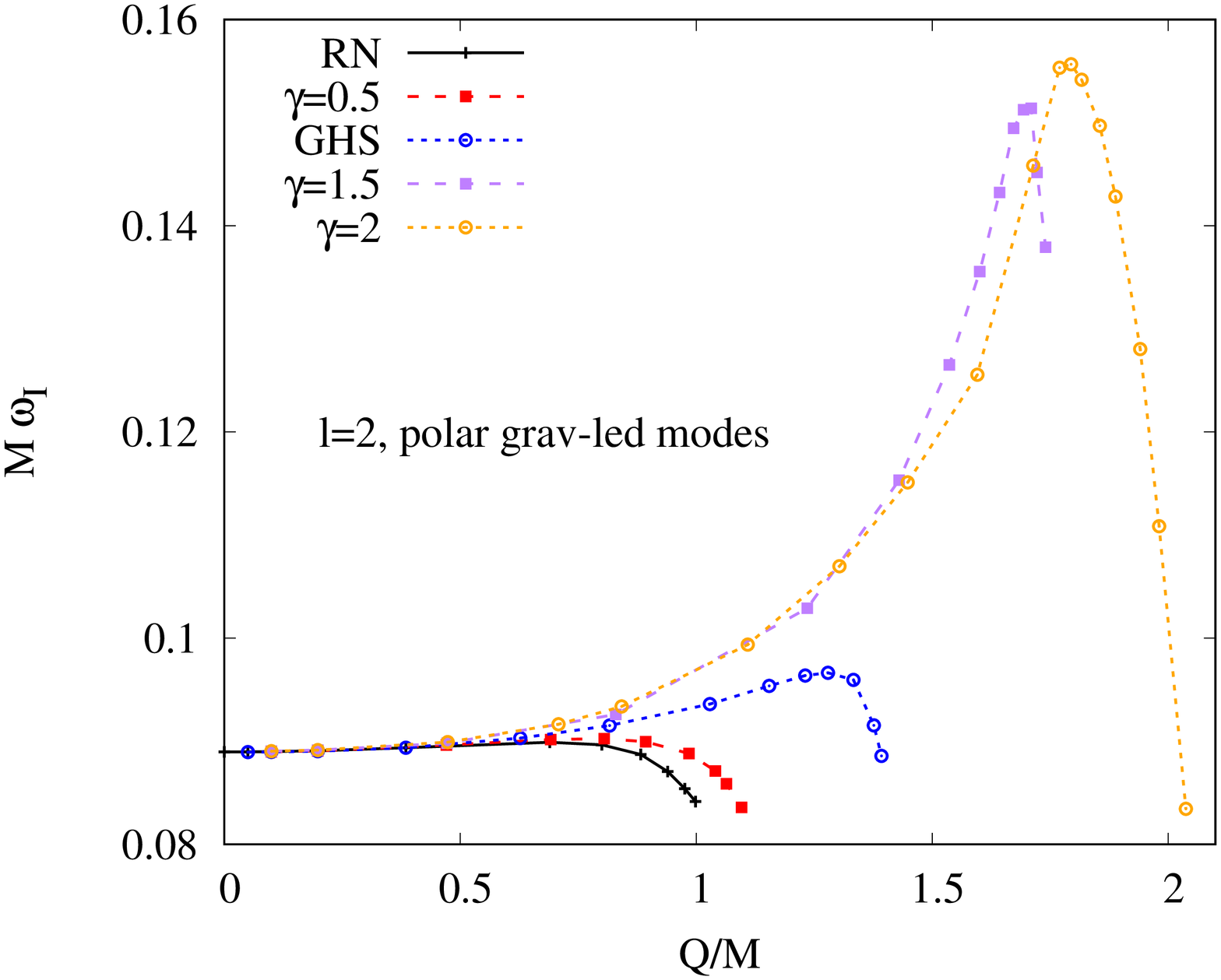}
	\caption{Grav-led modes for $l=2$ perturbations:
axial (upper) and polar (lower) modes.}
	\label{fig:l2_axial_polar_grav_R}
\end{figure}

Next we discuss the $l=1$ modes. 
These perturbations possess three families of modes: 
one with scalar-led modes, one with axial EM-led modes,
and one with polar EM-led modes.
In Fig.~\ref{fig:l1_polar_scalar_R}, we show the $l=1$ scalar-led modes, 
which appear when solving the polar perturbation equations. 
In Fig.~\ref{fig:l1_polar_scalar_R}(left), we see that the qualitative behavior 
of the real part $\omega_R$ is very similar to the $l=0$ scalar-led mode: 
the frequency grows when $Q/M$ and $\gamma$ increase, 
and tends to deviate strongly from the Schwarzschild
value close to the critical solutions. The imaginary part $\omega_I$, 
shown in Fig.~\ref{fig:l1_polar_scalar_R}(right), 
behaves somewhat differently from the case of the $l=0$ scalar-led modes, 
since now the steep rise in the vicinity of the maximal $Q/M$ is absent.
Consequently, the overall deviation from the GR values is not so large 
when compared with the previous case.

In Fig.~\ref{fig:l1_axial_polar_EM_R}, we show the $l=1$ EM-led modes 
for the axial (upper row) polar (lower row) perturbations.
The spectrum of RN EM-led modes coincides for both axial 
and polar perturbations. 
But as seen in the figures, when the coupling $\gamma$ is finite, 
the axial and polar modes no longer coincide. 
Comparison of the real part $\omega_R$ 
(Fig.~\ref{fig:l1_axial_polar_EM_R}(upper left) and 
Fig.~\ref{fig:l1_axial_polar_EM_R}(lower left)) 
reveals that for fixed values of $Q/M$ and $\gamma$, 
the polar modes are somewhat lower than the axial modes. 
This is also seen for the imaginary part $\omega_I$ 
(Fig.~\ref{fig:l1_axial_polar_EM_R}(upper right) 
and Fig.~\ref{fig:l1_axial_polar_EM_R}(lower right)). 
Nonetheless, the qualitative behavior of both axial and polar modes 
is rather similar.

\subsection{Spectrum of $l=2$ perturbations}

Last we consider the $l=2$ modes. 
These perturbations possess five families of modes: 
one of scalar-led modes, two of EM-led modes (axial and polar) 
and another two of grav-led modes (axial and polar).
In Fig.~\ref{fig:l2_polar_scalar_R}, we show the $l=2$ scalar-led modes. 
Again, they possess properties which are similar to the other 
scalar-led modes with lower $l$.

In Fig.~\ref{fig:l2_axial_polar_EM_R}, %
we show the EM-led modes for axial (upper row) 
and polar (lower row) perturbations.
Here the situation is qualitatively similar to $l=1$: 
while the RN modes possess isospectrality, 
this is broken when a non-trivial dilaton field is present.

In Fig.~\ref{fig:l2_axial_polar_grav_R}, %
we show the $l=2$ grav-led modes for axial (upper row) 
and polar (lower row) perturbations. 
As for the EM-led modes, both channels are degenerate for the RN modes, 
but isospectrality is broken for the charged black holes when $\gamma \neq 0$. 
Qualitatively both modes show a similar dependence on $Q/M$ and $\gamma$, 
with large deviations from GR arising close to the critical solution.

\section{Conclusions}

\begin{figure}[t]
	\centering
	\includegraphics[width=0.33\linewidth,angle=-90]{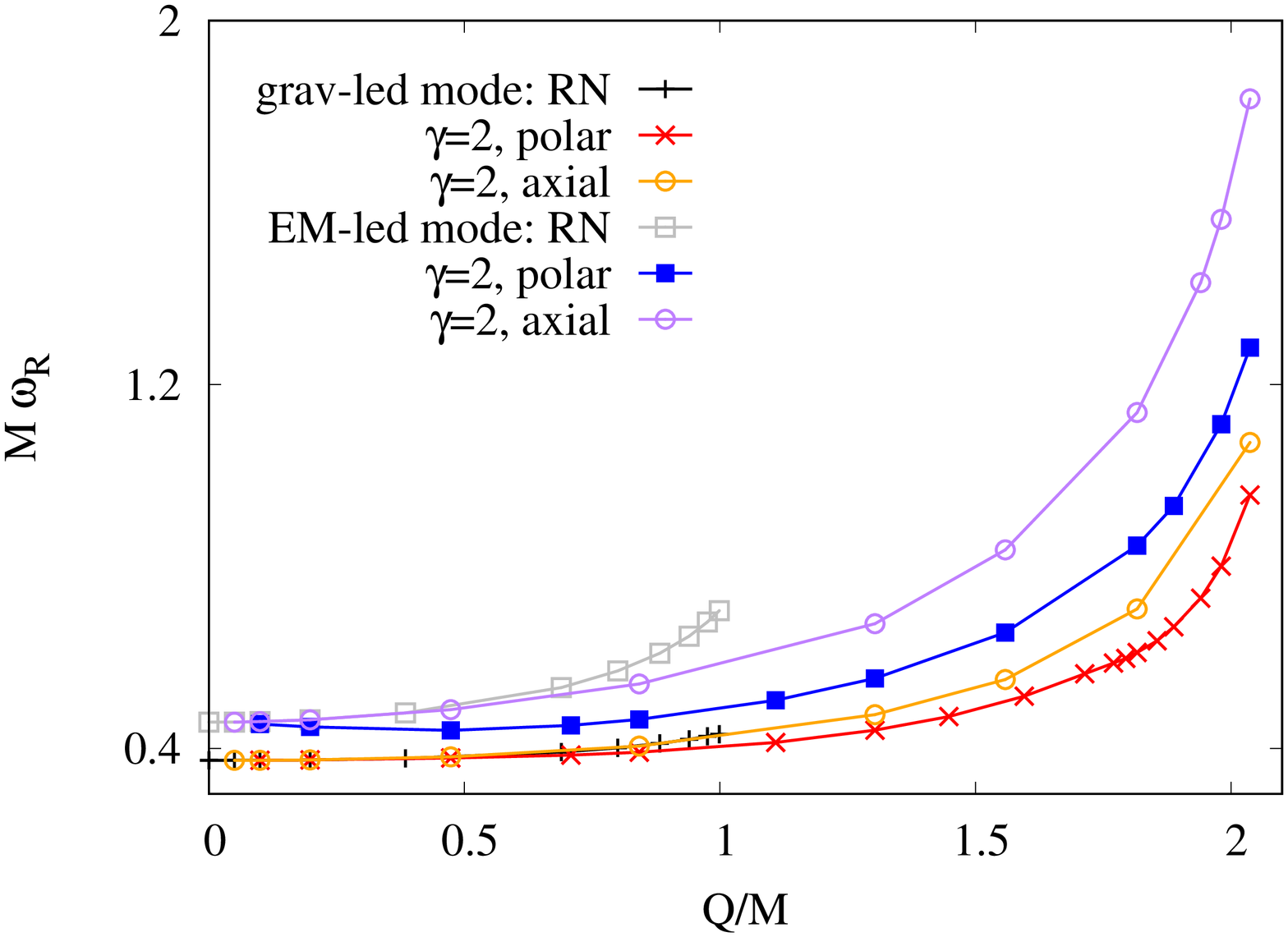}
	\includegraphics[width=0.33\linewidth,angle=-90]{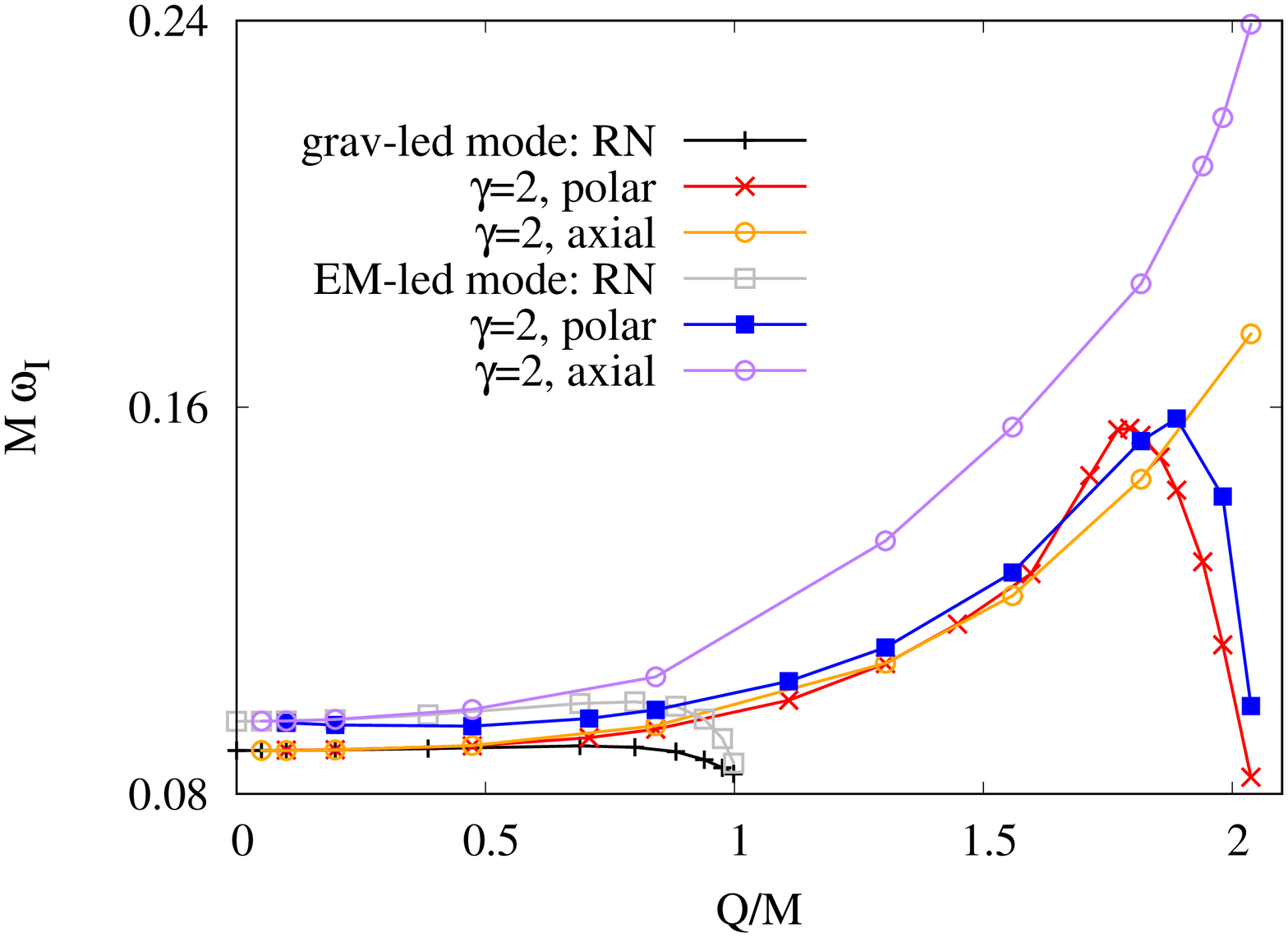}
	\caption{Illustration of broken isospectrality in the presence of a dilaton for $l=2$ modes.}
	\label{fig:iso_break}
\end{figure}

We have studied the QNMs of static spherically symmetric 
dilatonic electrically charged black holes,
considering the dilaton coupling constant $\gamma$ as a free parameter, 
and varying the charge within the allowed intervals, 
starting from the Schwarzschild solution 
all the way to the maximally charged black holes for a given $\gamma$,
extending previous work considerably 
\cite{Ferrari:2000ep,Konoplya:2001ji,Brito:2018hjh} 
(see also \cite{Hirschmann:2017psw} for dynamical evolution).
This allows us to conclude that these EMD black hole solutions are 
(mode-)stable under linear perturbations, 
since the imaginary part $\omega_I$ of the modes 
never changes sign. All the modes are damped modes.

In the presence of a non-trival dilaton background field, 
the perturbation equations 
also imply the presence of scalar radiation in a ringdown.
The analysis of the perturbation equations shows that in the polar case, all
the types of perturbations, scalar, vector and gravitational, are coupled,
whereas in the axial case, no scalar perturbations are present.
Thus for $l=2$, there arise five distinct modes: the grav-led (polar and axial),
the EM-led (polar and axial) and the scalar-led (polar) modes.
For $l=1$, there are no grav-led modes, 
and for $l=0$, there are only scalar-led modes.

Clearly, the presence of the scalar perturbations 
in the system of polar equations
and its absence in the system of axial equations 
makes these two systems distinctly different \cite{Ferrari:2000ep}. 
Therefore the breaking of the isospectrality of the RN and Schwarzschild
modes should not come as a surprise when a scalar background field is present.
Thus the simplicity of the spectrum of RN and Schwarzschild modes is lost,
and a much  richer spectrum appears.

This isospectrality breaking was already noted by 
Ferrari et al.~\cite{Ferrari:2000ep},
and investigated further by Pacilio and Brito \cite{Brito:2018hjh}, 
who, however, considered only small values of the electric charge. 
In their study they concluded that isospectrality breaking
is not so pronounced in the grav-led modes, 
but more prominent in the EM-led modes.
Moreover, for small charge, the grav-led modes 
do not exhibit much dependence on the dilaton coupling constant, 
while the dependence of the EM-led modes on $\gamma$ is stronger
\cite{Brito:2018hjh}.

By allowing for large values of the charge, 
we have shown that the situation changes.
We find large deviations from the Schwarzschild and RN modes,
that are typically the larger 
the larger the charge and the coupling constant.
The real parts $\omega_R$ of the modes always increase monotonically,
rising steeply for large charge and $\gamma$. 
The imaginary parts $\omega_I$, however, typically
change non-monotonically, but exhibit also steep rises or steep fall-offs
in the vicinity of the respective maximal charge.
Since these strong changes can be very different 
for the axial and the polar modes,
isospectrality becomes strongly broken, 
both for the EM-led and grav-led modes,
as illustrated in Fig.~\ref{fig:iso_break}.

The next step would be to investigate the QNMs of rotating EMD black holes
\cite{Frolov:1987rj,Rasheed:1995zv,Kleihaus:2003df}.
The QNMs of the Kerr-Newman black holes have been studied before
\cite{Pani:2013ija,Pani:2013wsa,Mark:2014aja,Dias:2015wqa}.
Pacilio and Brito \cite{Brito:2018hjh} have presented a first study
of the QNMs of slowly rotating EMD black holes for small values of
the charge. The challenge will be to extend these results
to large values of the charge, and, in particular, to fast rotation.

\begin{acknowledgments}
The authors gratefully acknowledge support by the
DFG Research Training Group 1620  \textit{Models of Gravity}
and the COST Action CA16104 \textit{GWverse}.
JLBS would like to acknowledge support from
the DFG Project No. BL 1553.
\end{acknowledgments}

\appendix*
\section{Perturbation equations}

Here we present a detailed discussion of the perturbation equations
for the axial and polar case.

\subsection{Axial case}

Inserting the ansatz (\ref{metric_axial_pert})
and (\ref{em_axial_pert}) into the field equations 
(\ref{Eins}), (\ref{Mxw}) and (\ref{Klein}), 
we obtain the following set of differential equations:
\begin{eqnarray}
\label{eqh0}
 \partial_r h_0 = - i\omega h_1
 -4e^{\gamma\phi_0}(\partial_r{a_{0}})W_2  +\frac{2}{r} h_{0}-{\frac {{i} }{2\omega\,{r}^{3}f  } h_{1}} \Big[ -\left( \partial_r\phi_{0}   \right) ^{2} f ^{2} {r}^{2}(r-2m) \nonumber \\-2rf^{2}l(l+1)
 +2rf\left( \partial_rf   \right)\left( r\partial_rm  + m - r \right) 
 -2{r}^{2}f\left( \partial_r^2 f   \right)(r-2m)   +4{r}^{2}{e^{\gamma\,\phi_{0}  }}f     \left( \partial_r{a_{0}}   \right) ^{2}(r-2m) \nonumber \\
  +\left( \partial_rf   \right) ^{2}  {r}^{2}(r-2m)+4\, \left( \partial_rm   \right) r f^{2}  +4f^2 (r-m) \Big] \ , \nonumber
\end{eqnarray}
\begin{eqnarray}
\partial_r h_1 =
{\frac {-i\omega r h_0 }{f  \left( r-2m \right) }} +h_1\frac{2f   \left( \partial_r m   \right) r-r\left( \partial_rf   \right) \left(r-2m\right) -2f  m}{2rf  \left( r-2m  \right) } \nonumber \ ,
\end{eqnarray}
\begin{eqnarray}
\partial_r^2{W_{2}}  =
{ \left(  \frac{4}{f}{e^{\gamma\phi_{0}}}\left(\partial_r a_{0}\right)^{2}-\frac{{\omega}^{2}{r}}{f(r-2m)}+\frac{l(l+1)}{r(r-2m)} \right) {W_{2}} }  -\left[ \frac{\partial_r f}{2f} - \frac{\partial_rm}{r-2m} + \gamma\partial_r\phi_0 + \frac{m}{r(r-2m)}  \right] {\partial_r{W_{2}}} \nonumber \\  +\frac{i(\partial_r a_0) h_2}{2\omega r^3 f^2} \Big[ 4f^2(r-m) + 4rf^2\partial_rm - r^2f^2(r-2m)(\partial_r\phi_0)^2 - 2rf^2l(l+1)
\nonumber \\ 
+4fr^2e^{\gamma\phi_0}(r-2m)(\partial_ra_0)^2 -2fr^2(r-2m)(\partial_r^2f)+r^2(r-2m)(\partial_rf)^2 + (2r^2f\partial_rm+2rf(m-r))\partial_rf\Big]
\nonumber \\ 
+h_0\Big[\Big(\frac{\partial_rf}{2f^2}+\frac{\partial_rm}{m(r-2m)}-\frac{\gamma\partial_r\phi_0}{f}+\frac{3m-2r}{fr(r-2m)}\Big)\partial_ra_0-\frac{\partial_r^2a_0}{f}\Big] \ . 
\end{eqnarray}

This is a system of coupled differential equations: 
two first order differential equations for $h_0$ and $h_1$ 
are coupled with a second order differential equation for $W_2$. 

In order to obtain the QNMs of the axial perturbations, 
we need to impose proper boundary conditions.
Imposing the outgoing wave behavior,
this implies that at infinity, 
the perturbation functions behave like
\begin{eqnarray}
h_0 = r e^{i\omega R}  \left[A_g^+\left(-\omega+O(r^{-1})\right)+A_{F}^+\left(\frac{Q}{r^2}+O(r^{-3})\right)\right] \ ,  \nonumber \\
h_1 = r\omega e^{i\omega R} \left[A_g^+\left(1+O(r^{-1})\right)+A_{F}^+\left(\frac{-Q}{\omega r^2}+O(r^{-3})\right)\right] \ ,  \nonumber \\
W_2 = e^{i\omega R} \left[A_g^+\left(\frac{Q(l-1)(l+2)}{4\omega r^2}+O(r^{-3})\right)+A_{F}^+\left(1+O(r^{-1})\right)\right] \ .
\label{axout}
\end{eqnarray}

Close to the horizon, the perturbations have to be ingoing, 
implying that
\begin{eqnarray}
h_0 &=& e^{-i\omega R}\left[A_g^-\left(1+O(r-r_H)\right) + A_{F}^-\left(O(r-r_H)\right)\right] \ ,  \nonumber \\
h_1 &=& \frac{\omega}{r-r_H}e^{-i\omega R}\left[A_g^-\left(\frac{r_H^{3/2}e^{\gamma\phi_H/2}}{\omega\sqrt{f_1\left(r_H^2e^{\gamma\phi_H}-Q^2\right)}}+O(r-r_H)\right) + A_{F}^-\left(O(r-r_H)\right)\right] \ , \nonumber \\
W_2 &=& e^{-i\omega R}\left[A_g^-\left(O(r-r_H)\right)+A_{F}^-\left(1+O(r-r_H)\right)\right] \ . 
\label{axin}
\end{eqnarray}

Note that the expansion is now characterized by two amplitudes, 
the space-time perturbation amplitude $A_g^{\pm}$ 
and the electromagnetic perturbation amplitude $A_{F}^{\pm}$.

\subsection{Polar case}

Analog to the axial case,
we can plug the ansatz for the polar perturbations
(\ref{metric_polar_pert}), (\ref{A_polar}) and (\ref{s_polar})
into the field equations (\ref{Eins}), (\ref{Mxw}) and (\ref{Klein}). 
This results in the following set of equations: 
\begin{eqnarray}
L+N=0 \ , \nonumber
\end{eqnarray}

\begin{eqnarray}
\partial_rH_1  =4{e^{\gamma\phi_0}} \left( \partial_ra_0\right) (W_1-\partial_rV_1)  +\frac{\left(\partial_r\delta		  \right)r(r-2m)+2\left(\partial_rm   \right)r-2 m
}{r \left( r-2 m  \right)} H_1 
-\frac{2 i\omega r^{2}}{r \left( r-2 m  \right)}(T+L) \ , \nonumber
\end{eqnarray}

\begin{eqnarray}
\partial_r(N + T)  = {\frac {{r}^{3} \left( \partial_r \phi_0   \right) ^{2}(r-2m)-4{Q}^{2}e^{-\gamma\phi_{0}}   }{8	\left( r-2m   \right) {r}^{2}} (L-N)} 
+\frac{r-3m}{r(r-2m)}N
\nonumber \\
+\frac{r-m}{r(r-2m)}L
- \frac{2e^{2\delta}}{r(r-2m)}\left(Q(i\omega V_{1}+a_{1})-\frac{i}{4}\omega r^{2}H_{1}\right)+\frac{1}{2}\phi_1  (\partial_r \phi_0) \ , \nonumber
\end{eqnarray}

\begin{eqnarray}
\partial_rN =-\frac{r}{4} \left( \partial_r \phi_0   \right) (\partial_r\phi_{1}) 
- \frac{1}{r-2m} \Big[ \Big( -\left( \partial_r\delta   \right)r(r-2m)
- \left( \partial_rm \right) r-m  +r \Big) (\partial_rT)   -{e^{2\delta+\gamma \phi_{0}}{r}^{2}\left(\partial_ra_0   \right)(\partial_ra_1)}
 \nonumber \\
-{i e^{2\delta}\omega rH_1}-L-\frac {1}{2} \Big(2{e^{2\delta+\gamma\phi_0}}{r}^{2} \left(\partial_ra_0\right)^{2}+l(l+1) \Big) N
-{\frac {1}{2} {e^{2\delta+\gamma\phi_{0}}\gamma\left(\partial_ra_0\right)^{2}{r}^{2}\phi_1}} 
\nonumber \\
+{\frac{2{r}^{3}e^{2\delta}{\omega}^{2}-l(l+1)(r-2m)+2(r-2m)}{2\left(r-2m\right)}T  }-i \left( \partial_ra_0   \right) e^{2\delta+\gamma\phi_{0}}		\omega\,{r}^{2}W_1\Big] \ , \nonumber
\end{eqnarray}

\begin{eqnarray}
\partial_rT  ={\frac {i}{8\omega r^{2}} \left( 4{e^{2\delta+\gamma\phi_0 }}{r}^{2} \left( \partial_r a_0 \right)^{2}+r\left( \partial_r \phi_0 \right) ^{2}(r-2m)+2{l}(l+1)-8(\partial_rm) \right) H_1 }
\nonumber \\
+{\frac {L}{r}}+\frac{1}{4}\phi_1  (\partial_r \phi_0)-{\frac{1}{r\left( r-2m\right)}\Big[r\left(\partial_r\delta   \right) (r-2m)+r((\partial_rm) +1) -3m\Big] T} \ , \nonumber
\end{eqnarray}

\begin{eqnarray}
\partial_r^2T  ={\frac {1}{r}}(\partial_rL)+ \frac{1}{4}\, \left( \partial_r \phi_0   \right) (\partial_r\phi_1)+\frac{1}{4r(r-2m)}\Big[ 4\left( 		r\left( \partial_rm   \right) +5\,m		 -3\,r \right) (\partial_rT) + {4 \left( \partial_ra_0   \right)   {e^			{2\delta+\gamma\phi_{0}  }}r^2(\partial_ra_1)  } \nonumber \\
+2{r^{2}  {e^{2\delta+\gamma\phi_{0}  }}	\gamma\left(\partial_ra_0   \right)^{2}\phi_1 }+{ \left( 4\,{			e^{2\delta+\gamma\phi_0}}{r}^{2} \left( \partial_ra_0   \right) ^{2}+r \left( \partial_r \phi_0   \right) ^{2}(r-2m)+2l(l+1)-4\,(2(\partial_rm)  		-1) \right) L  }
\nonumber \\
-{ \left(r \left( \partial_r		 \phi_0   \right) ^{2}(r-2m)  -8		\,\partial_rm   \right) N }+2\,{		\left( l+2 \right)  \left( l-1 \right) T  }+{4i \left( \partial_ra_0   \right)   {e^{2\delta+\gamma\phi_{0}				 }} {\omega\,r^{2}W_1}}\Big] \ , \nonumber
\end{eqnarray}

\begin{eqnarray}
\partial_r^2(T+N)=
\frac{i\omega r e^{2\delta}}{r-2m}(\partial_rH_1)
-\frac{2-3r(\partial_r\delta)-4(\partial_rm)+6m(\partial_r\delta)}{r-2m}(\partial_rN) 
+ \frac{1}{2}(\partial_r\phi_0)(\partial_r\phi_1)
\nonumber \\
+ \frac{r(r-2m)(\partial_r\delta)+2m-2r+2r(\partial_rm)}{r-2m}(\partial_rT)
- \frac{2re^{\gamma\phi_0+2\delta}}{r-2m}(\partial_ra_0)(\partial_ra_1) 
- \frac{i\omega e^{2\delta}(m-r+r\partial_rm)}{(r-2m)^2}H_1
\nonumber \\
-\frac{2ri\omega e^{\gamma\phi_0+2\delta}(\partial_ra_0)}{r-2m}W_1
-\frac{r\gamma e^{\gamma\phi_0+2\delta}(\partial_ra_0)^2}{r-2m}\phi_1
+\frac{e^{2\delta}\omega^2 r^2}{(r-2m)^2}(N-T)+\frac{2re^{\gamma\phi_0+2\delta}(\partial_ra_0)^2}{r-2m}T 
\nonumber \\
+\Big[
2\partial_r^2\delta+\frac{2}{r-2m}\partial_r^2m-\frac{1}{2}(\partial_r\phi_0)^2-2(\partial_r\delta)^2+\frac{2(\partial_r\delta)(m+r-3r(\partial_rm))}{r(r-2m)}
\Big](N+T) \ , \nonumber
\end{eqnarray}

\begin{eqnarray}
\partial_rV_1  =\frac {i{e^{2\delta}}\omega{r}^{3}}{l(l+1)(r-2m)}  \Big[(\partial_ra_1)  +\left( \partial_ra_0   \right)(L+N-2T+\gamma\phi_1)\Big] + \Big[1-\frac{\omega^2r^3e^{2\delta}}{l(l+1)(r-2m)}\Big]W_1 \ , \nonumber
\end{eqnarray}
\begin{eqnarray}
\partial_r^2V_1  = 
-{\frac{{r}^{2} e^{2\delta}{\omega}}{(r-2m)^{2}} \Big[\omega V_1-ia_1\Big]} 
+\partial_rW_1
\nonumber \\ 
-\frac {1}{r(r-2m)}  \Big[2(m-r\partial_r m)+r(r-2m)\big(\gamma(\partial_r \phi_0)-(\partial_r \delta)\big)  \Big] [\partial_rV_1-W_1] \ , \nonumber
\end{eqnarray}

\begin{eqnarray}
\partial_r^2a_1  =  -i\omega(\partial_rW_1)  - \left(\partial_ra_0\right) \big[\partial_r(L+N-2T+\gamma\phi_1)\big]  
 \nonumber \\
 +{\frac{l(l+1)}{r \left( r-2m \right) }}(i\omega V_1+a_1)-\left(\gamma\left( \partial_r \phi_0\right) +(\partial_r\delta)+\frac{2}{r} \right)\big[(\partial_r a_1)+i\omega W_1\big]
\nonumber \\
-{\Big[\left( \partial_r a_0   \right)  \left( \partial_r \phi_0   \right) \gamma+ \left( \partial_r \delta   \right) (\partial_r a_0)  +(\partial_r^2a_0)  +\frac {2}{r}(\partial_ra_0)\Big] [\gamma\,\phi_1+2(L+N)] } \ , \nonumber
\end{eqnarray}

\begin{eqnarray}
\partial_r^2\phi_1  =\left( \partial_r \phi_0   \right) \Big[\partial_r(N+2T-L)\Big]
-\frac{ir\omega e^{2\delta}\partial_r\phi_0}{r-2m}H_1-\frac{4r\gamma e^{\gamma\phi_0+2\delta}}{r-2m}N-\frac{4ri\omega\gamma e^{\gamma\phi_0+2\delta}}{r-2m}W_1 \nonumber \\
-\frac{4r\gamma e^{\gamma\phi_0+2\delta} (\partial_ra_0)}{r-2m} (\partial_ra_1)+\frac{2r(\partial_rm)+2m-2r+r(r-2m)(\partial_r\delta)}{r(r-2m)}(\partial_r\phi_1) \nonumber \\
+\Big[2\partial_r\phi_0\partial_r\delta-2\partial_r^2\phi_0-\frac{4r\gamma e^{\gamma\phi_0+2\delta}(\partial_ra_0)^2}{r-2m}+\frac{4(\partial_r\phi_0)(m-r+r\partial_rm)}{r(r-2m)}\Big]L \nonumber \\
+\Big[\frac{l(l+1)}{r(r-2m)}-\frac{r^2\omega^2e^{2\delta}}{(r-2m)^2}-\frac{2r\gamma^2e^{\gamma\phi_0+2\delta}(\partial_ra_0)^2}{r-2m}\Big]\phi_1 \ .
\end{eqnarray}

This system of equations can be simplified. With the redefinitions
\begin{eqnarray}
\label{redef}
F_0(r) &=& -i\omega W_1(r) - \frac{dW_1(r)}{dr} \ , \\
F_1(r) &=& -i\omega V_1(r) - a_1(r) \ , \\
F_2(r) &=& -W_1(r) + \frac{dV_1(r)}{dr} \ ,
\end{eqnarray}
the minimal set of differential equations 
(Master equations) can again be written in vectorial form,
but now in terms of a complicated $6\times 6$ matrix. 

The QNMs of the polar perturbations are again obtained 
imposing the outgoing wave behavior at infinity.  
The perturbation functions behave like
\begin{eqnarray}
H_1 = r e^{i\omega R}  \left[A_g^+\left(-2+O(r^{-1})\right)+A_{F}^+\left(\frac{2iQ}{\omega r^2}+O(r^{-3})\right)+A_{\phi}^+\left(\frac{-i}{4\omega^2 r^4}(\omega\gamma Q^2-iQ_S)+O(r^{-5})\right)\right] \ ,  \nonumber \\
T = e^{i\omega R}  \left[A_g^+\left(1+O(r^{-1})\right)+A_{F}^+\left(\frac{-iQ}{\omega r^2}+O(r^{-3})\right)+A_{\phi}^+\left(\frac{iQ_S}{4\omega r^3}+O(r^{-4})\right)\right] \ ,  \nonumber \\
F_0 = \frac{1}{r^2} e^{i\omega R}  \left[A_g^+\left(-2Q+O(r^{-1})\right)+A_{F}^+\left(\frac{-il(l+1)}{\omega}+O(r^{-1})\right)+A_{\phi}^+\left(\frac{\gamma Q}{r}+O(r^{-2})\right)\right] \ ,  \nonumber \\
F_1 = e^{i\omega R}  \left[A_g^+\left(\frac{Q}{r}+O(r^{-2})\right)+A_{F}^+\left(1+O(r^{-1})\right)+A_{\phi}^+\left(\frac{-\gamma Q}{4r^2}+O(r^{-3})\right)\right] \ ,  \nonumber \\
\phi_1 = \frac{1}{r}e^{i\omega R}  \left[A_g^+\left(\frac{2Q_SM-2\gamma Q^2}{r}+O(r^{-1})\right)+A_{F}^+\left(\frac{Q(\gamma l(l+1)-i\omega Q_S)}{\omega^2 r^2}+O(r^{-3})\right)+A_{\phi}^+\left(1+O(r^{-1})\right)\right] \ . 
\label{polout}
\end{eqnarray}

Close to the horizon, the perturbation has to travel as an ingoing wave, 
meaning
\begin{eqnarray}
H_1 = \frac{1}{r-r_H} e^{-i\omega R}  \left[A_g^-\left(
\frac{2ir_H^2(2ie^{\gamma\phi_H+\delta_H}\omega r_H^3+e^{\gamma\phi_H} r_H^2-Q^2)}{(2r_H e^{\delta_H}\omega+il(l+1))(e^{\gamma\phi_H}r_H^2-Q^2)} + O(r-r_H)\right)
\right. \nonumber \\ \left.
+A_{F}^-\left(\frac{	4e^{\delta_H+\gamma\phi_H}Q r_H^3}{(r_H^2e^{\gamma\phi_H}-Q^2)l(l+1)}(r-r_H)+O((r-r_H)^2)\right)
\right. \nonumber \\ \left.
+A_{\phi}^-\left(\frac{-4\gamma r_H Q^2 r_H^3}{(r_H^2e^{\gamma\phi_H}-Q^2)l(l+1)}(r-r_H)+O((r-r_H)^2)\right)\right] \ ,  \nonumber \\
T = e^{-i\omega R}  \left[A_g^-\left(
1 + O(r-r_H)\right)
\right. \ \ \ \ \ \ \ \ \ \ \ \ \ \ \ \ \ \ \ \ \ \ \ \ \ \ \ \ \ \ \ \ \ \ \ \ \ \ \  \ \ \ \ \ \ \ \ \ \ \ \ \ \ \ \ \ \ \ \nonumber \\ \left.
+A_{F}^-\left(\frac{-2e^{\delta_H+\gamma\phi_H}Q r_H(2r_H e^{\delta_H}\omega+il(l+1))}{(2r_H^3e^{\gamma\phi_H+\delta_H}\omega+ie^{\gamma\phi_H}r_H^2-iQ^2)l(l+1)}(r-r_H)+O((r-r_H)^2)\right)
\right. \nonumber \\ \left.
+A_{\phi}^-\left(\frac{-\gamma Q^2(2r_H e^{\delta_H}\omega+il(l+1))}{(2r_H^3e^{\gamma\phi_H+\delta_H}\omega+ie^{\gamma\phi_H}r_H^2-iQ^2)l(l+1)}(r-r_H)+O((r-r_H)^2)\right)\right] \ ,  \nonumber \\
F_0 = e^{-i\omega R}  \left[A_g^-\left(O(r-rH)\right)+A_{F}^-\left(1 + O(r-r_H)\right)+A_{\phi}^-\left(O(r-rH)\right)\right] \ ,  \nonumber \\
F_1 = e^{-i\omega R}  \left[A_g^-\left(\frac{-2i\omega e^{-\gamma\phi_H}}{l(l+1)}+O(r-r_H))\right)+A_{F}^-\left(\frac{-ie^{\delta_H}\omega r_H^2}{l(l+1)}+O(r-rH)\right)
\right. \nonumber \\ \left.
+A_{\phi}^-\left(\frac{i\gamma\omega Q e^{-\gamma\phi_H}}{l(l+1)}+O(r-rH)\right)\right] \ ,  \nonumber \\
\phi_1 = e^{-i\omega R}  \left[A_g^-\left(O(r-rH)\right)+A_{F}^-\left(O(r-rH)\right)+A_{\phi}^-\left(1 + O(r-r_H)\right)\right] \ . 
\label{polin}
\end{eqnarray}

In the previous expressions we have three undetermined amplitudes:  
the space-time perturbation amplitude $A_g^{\pm}$, 
the electromagnetic perturbation amplitude $A_{F}^{\pm}$ 
and the scalar perturbation amplitude $A_{\phi}^{\pm}$.

\bibliography{paper_dbh_bib}

\end{document}